\documentclass[a4paper,11pt,openany]{book}
\usepackage[applemac]{inputenc} 
\usepackage[T1]{fontenc} 
\usepackage{geometry} 
\usepackage{amssymb,amsmath} 
\usepackage{graphicx} 
\usepackage[francais,english]{babel} 
\usepackage{mathrsfs} 
\usepackage{hyperref}
\usepackage{cite}
\usepackage{color}
\usepackage{pdfcomment}

\newcommand{\kB}{\ensuremath{k_{\mathrm{B}}}} 
\newcommand{\ave}[1]{\ensuremath{\left \langle {#1} \right \rangle}} 
\newcommand{\eps}{\ensuremath{\varepsilon}} 

{\large }
\newcommand{\om}{\ensuremath{\omega}}
\newcommand{\Om}{\ensuremath{\Omega}}

\newcommand{\p}{\left(}
\newcommand{\q}{\right)}

\newcommand{\g}[1]{\og{#1}\fg}
\newcommand{\fr}[1]{\foreignlanguage{francais}{#1}}

\providecommand{\HUGE}{\Huge}
\newlength{\drop}
\newcommand*{\titleGM}{\begingroup
\drop=0.1\textheight
	\hbox{%
	\hspace*{0.2\textwidth}%
	\rule{1pt}{\textheight}
	\hspace*{0.05\textwidth}%
	\parbox[b]{0.75\textwidth}{ \vbox{%
		\vspace{\drop}
		{\noindent\HUGE\bfseries Democritus}\\[2\baselineskip]
		{\LARGE\bfseries and the motive power of fire}\\[4\baselineskip] 
		{\Large Jacques ARNAUD}\par
		{\itshape Mas Liron, F30440 Saint Martial, France}\\[1\baselineskip]
		{\Large Laurent CHUSSEAU}\par
		{\itshape IES, Université Montpellier II, F34095 Montpellier, France}\\[1\baselineskip]
		{\Large Fabrice PHILIPPE}\par
		{\itshape LIRMM, Université Montpellier II, F34392 Montpellier, France}\par
		\vspace{0.4\textheight}
		{\noindent \today}\\[\baselineskip] 
		}
		}
	}
\vfill
\null
\endgroup}

\begin{document}
\frontmatter
%
\thispagestyle{empty}
\titleGM
\clearpage
%
%

\section*{Abstract}

The present work is a translation from french to english of our previous \g{Démocrite et la puissance motrice du feu}, amended on a number of respects. It is mainly of historical and pedagogical interest. We suggest that the concepts introduced in the ancien Greece by Anaximander (flat earth) and Democritus (corpuscles moving in vacuum) allow us to obtain through qualitative observations and plausible generalizations the maximum efficiency and work of heat engines, results that were firmly established around 1824 by Carnot. A prologue introduces the subject. We next present the concept of thermal equilibrium and consider a model consisting of two reservoirs located at different altitudes, each with $g$ sites. Each site may contain a specified number of corpuscles. One particular site plays the role of \g{working agent}. We subsequently consider an alternative model consisting of independent corpuscles submitted to gravity and in contact with heat baths. Only average quantities are considered, leaving out fluctuations and questions of stability.

\begin{center}
\includegraphics[width=0.7\columnwidth]{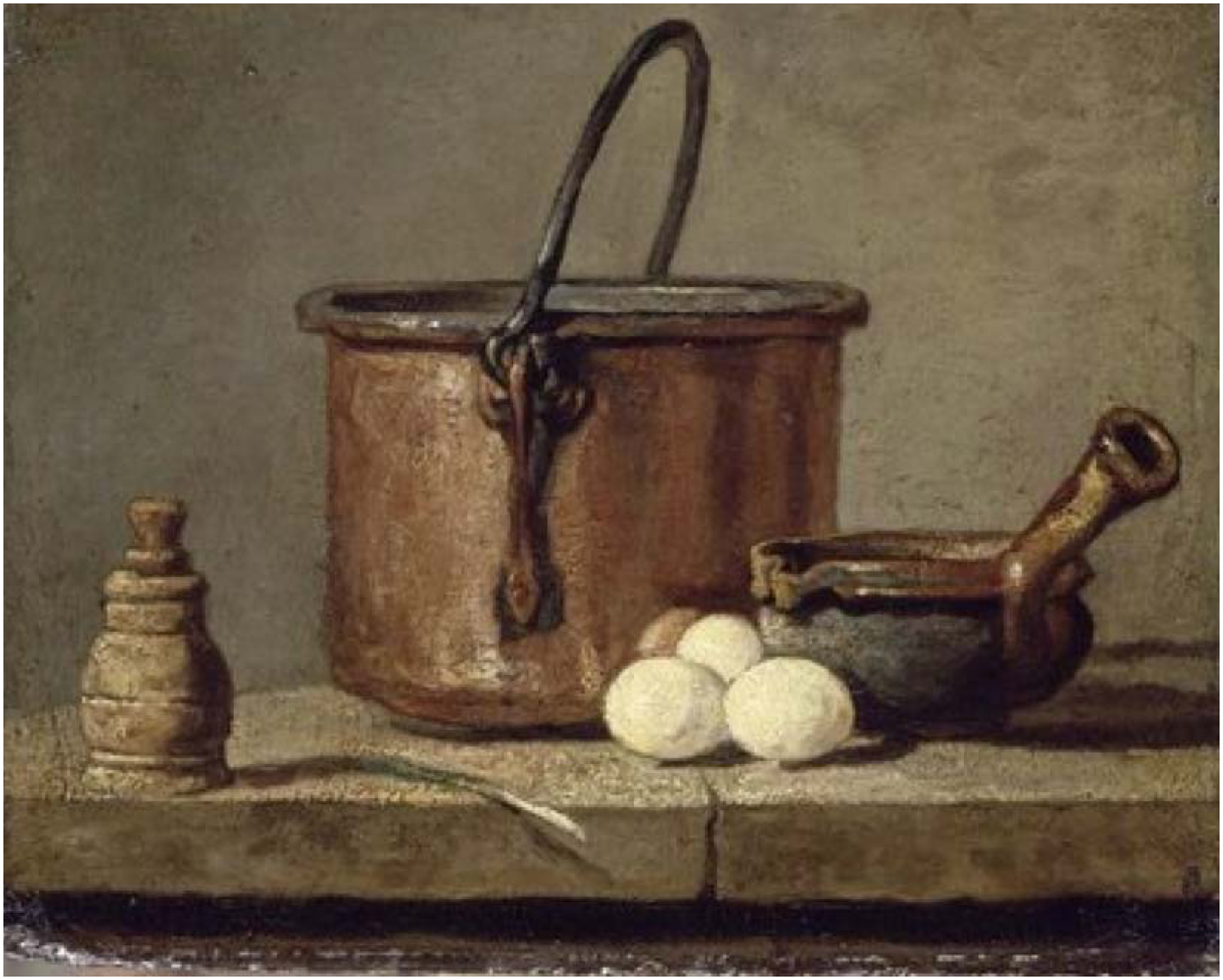}
\vspace{2em}
\begin{quote}
This painting by Chardin illustrates the \g{parti pris des choses} (Francis Ponge) adopted by Anaximander and Democritus in the ancient Greece. Aristotle notes that \g{Democritus omits to consider final causes and thus brings back to necessity every action of nature.} One can fancy that the Chardin painting represents three corpuscles and two sites, one being able to contain only few corpuscles, the other being able to contain many.
\end{quote}
\label{chardin}
\end{center}
\vfill~
\newpage
\tableofcontents

\mainmatter
\chapter{Introduction}\label{introduction}

Because the french version of this work \cite{ARNAUD:2011:HAL-00583100:1} attracted some attention \cite[see Ref. 31]{Abreu:2012} we are encouraged to present an english version of that earlier work, improved on some respects. It can be considered an introduction to heat laws. We are thinking primarily of readers who have no prior knowledge in Mathematics or Physics, but are willing to memorize a number of observations. Those readers may wish to consult easy-reading mainly historical books, such as \cite{Cardwell:1971}. A critical review of some modern concepts is given in \cite{Lavenda:2010}. Books relating to general concepts concerning waves, coherent and incoherent, may be useful, e.g. \cite{Arnaud:1976}.

After recalling laws of thermodynamics that do not involve microscopic considerations but involve some empirical facts, essentially as they were discovered by Carnot in 1824, we present two simple but unconventional approaches to the thermodynamics of ideal gases (independent corpuscles) that seem to be novel and lead to generally known results. Absolute temperatures are defined, as is usual since Kelvin time, such that the maximum heat-engine efficiency be $ 1-\text{lower absolute temperature}/\text{higher absolute temperature}$. Real gases, phase changes, the approach to equilibrium and fluctuations are not considered in this work.

In the first approach we consider reservoirs consisting of sites containing corpuscles, located at two different altitudes. A cycle consists of exchanging a site between the upper and lower reservoirs. In the second approach, we consider corpuscular motion under gravity such that the ideal-gas law be obeyed independently of the laws of motion (principle of simplicity). 

The theory presented is classical in the sense that the Planck constant $\hbar$ is arbitrarily small but we do not suppose that the corpuscle speeds are much smaller than the speed of light $c$ in free space, except in examples. Nor do we postulate any particular method of synchronization of distant clocks. It seems to us that the usual presentations of thermodynamics involve theoretical considerations and empirical results that may be superfluous. Democritus philosophy ($\sim$ -300) suffices to establish the ideal-gas law and the barometric law. These laws were verified at the time of Boyle ($\sim$ 1700) but they could have been verified at the time of Democritus because no sophisticated technology is needed. The macroscopic theory of heat engines was given by Carnot ($\sim$ 1824). Boltzmann ($\sim$ 1860) reintroduced discussions based on corpuscles, in agreement with the ancient-Greeks considerations.

Let us first quote Feynman \cite{Feyman:1977}: ``If, in some cataclysm, all of scientific knowledge were to be destroyed, and only one sentence passed to the next generations of creatures, what statement would contain the most information in the fewest words? I believe it is the hypothesis that all things are made of atoms. In that one sentence there is an enormous amount of information about the world, if just a little imagination and thinking are applied''. 

In line with this quotation, it seems worthwhile exploring how much physics may be derived from the corpuscular concept. Only some qualitative observations are needed.  A first observation is that two bodies left in contact for a sufficient period of time tend to reach the same temperature, as one can judge by our senses. Note that our senses may be misleading on that respect: if a piece of copper and a piece of wool are both at room temperature one gets the feeling that the piece of copper is colder than the piece of wool. This is because copper has a greater heat conductivity than wool, and thus takes more quickly heat out of our body (room temperature may be 20 °C while body temperature may be 37 °C). This example teaches us that, even though observations are the ultimate judges for the validity of a physical theory, besides consistency, our senses may not be thrust entirely. This warning was made by Democritus. 

The second observation is that it takes some time, denoted here $\tau(z_m)$, for a corpuscle thrown upward on earth to reach an altitude $z_m$ and come back to the ground level. We have of course $\tau(0)=0$. If the corpuscle bounces elastically on the ground, $\tau(z_m)$ represents the oscillation period. Under our assumptions the $\tau$-function does not depend on the initial altitude or initial time. Thus the time during which the corpuscle is located above some altitude $z\le z_m$ during a period is: $\tau(z_m-z)$. Our principle of simplicity then shows that the distribution of corpuscle energy necessarily involves a quantity with the dimension of energy that we denote: $\theta$. We prove that $\theta$ is an absolute temperature. 

Note that according to the definition given above (from the Carnot efficiency), absolute temperatures are defined only to within an arbitrary multiplicative factor. This factor is fixed by convening that $\theta$ equals some specific energy (in joules) at the water triple point, at which the solid, liquid and gas forms of water (or some other substance such as hydrogen) are in equilibrium. 

Let us cite other fairly common observations. If we obturate the outlet of pumps of the kind used to inflate tires, one can make the following observations: First, in order to compress air one must exert an increasing force on the piston and, on that respect, air seems to behave as a spring. Second, one feels that the air gets warmer (this is unlike the case of a spring that stores energy in an orderly manner rather than converting it into  heat). Third, if we wait for a few minutes the compressed air temperature becomes nearly equal to the room temperature, and one feels that the force that must be exerted to maintain the piston at the same position becomes smaller (but does not vanish). Fourth, if we then pull out the piston, one finds that the air temperature becomes \emph{smaller} than the room temperature, at least for a while. Anticipating the detailed laws given later on in this paper, let us say that, initially, the pump cylinder (with unit cross-section area and piston height $h$) is submitted to the atmospheric force. We initially take the atmospheric force as our force unit. It is balanced by the cylinder air force. When we push the piston fairly rapidly we are in the so-called adiabatic regime. The precise expression of the force in that regime will be given. It involves the notion of \g{internal energy}.

The main purpose of thermodynamics is to explain such observations in a precise manner, and design cyclic machines that generate work out of heat, or else cool out substances, as efficiently as possible. We now turn to historical and philosophical considerations.

\begin{figure}
\centering
\includegraphics[width=0.7\columnwidth]{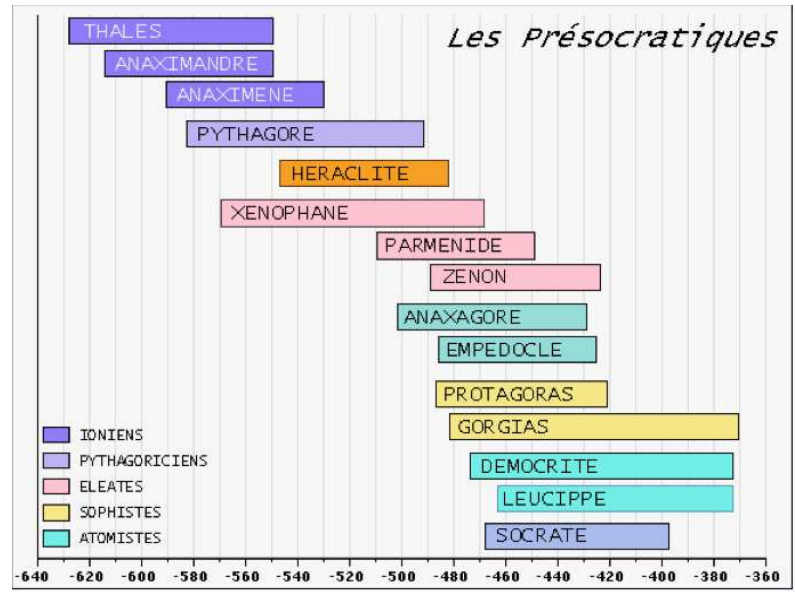}
\caption{This figure gives a list of the most important pre-socratic philosophers, see \url{http://www.igor-brevnjovski.net/}. The contributions of these scientists are described in: \url{http://coll-ferry-montlucon.planet-allier.com/gdscient.htm}.}
\label{presocratiques}
\end{figure}

The role of human beings in the universe has been well understood by pre-socratic Greek philosophers (for a list of the most important of these philosophers, see Fig.~\ref{presocratiques}). They anticipated up to a point the theory of evolution, considering that man is a branch of a tree emerging from duplication and selection. In particular, Anaximander asserted that men were originally fishes that loosed their scales in climbing on earth. The concept of transformation prominent in pre-socratic thinking was emphasized in the modern time through the detailed and cogent observations made by C. Darwin. Objections were raised against Darwin theory of evolution. In particular, Kelvin calculated the age of earth considering the heat carried in by the sun, the energy freed by the earth contraction, and the radiation of heat to the outer space. In that manner, he obtained an age of about one million years, which was insufficient for the evolution to have taken place. Kelvin at that time was unaware of the radioactivity of earth (uranium and thorium) that supplies an extra heat-source. Additionally, the temperature gradient at the earth surface is not representative of the temperature distribution deep into the earth. Present calculations give for the earth age about four billion years, with life appearing about three and an half billion years ago. Secondly, life generates order and thus reduces the entropy (to be defined later on), apparently contradicting the second law of Thermodynamics. However, natural sources of entropy (e.g., the heat from the sun) are immensely larger. The discovery of genes, with only some of them coding for proteins, led to minor revisions of the Darwin theory. In consideration of the title of this paper, it is appropriate to recall that Heraclitus considered fire as being the fundamental element.

The materialistic view-point is the radical originality of many pre-socratic thinkers (Democritus is considered a pre-socratic thinker even though he lived at the same time as Socratus \cite{rovelli}), some considering as a fundamental element the fire, the water, the air, or four elements, perhaps five. The Democritus theory distinguishes itself from these view-points by its coherence and simplicity. Detailed presentations of Democritus concepts may be found in books by Zeller \cite{zeller},  Salem \cite{salem2} or an article by Horne \cite{horne}, for example. The word \g{materialism} employed above may be confused with a kind of selfish social behavior. Furthermore this word may erroneously suggest that a fundamental status is ascribed to matter. Thus it would be best to employ instead the word: \g{emergentism}. Emergentism is the belief that thermal phenomena \g{emerge} from corpuscular motions, that life emerges from the laws of chemistry, and so on (see later on in this introduction). It is justified by its power of prediction. Some laws are preserved in the process of emergence, for example the law of energy conservation in isolated systems. Furthermore, there may be at one stage a remembrance of the previous stages, usually expressed in terms of \emph{fluctuations}. We gave in \cite{arnaud4} a theory of laser fluctuations related to the present considerations. 

It must be admitted that the emergentism view-point has loose ends. One is that the standard model of particles (including the recently discovered Higgs field) does not yet explain the origin of the symmetries found in nature and, on the other hand, gravity remains poorly understood. Second, even through neurosciences are able to relate our desires to hormones such as the ocytocine, the nature of thoughts as perceived by us may perhaps be discussed only through art and ethic (strangeness, charm, beauty and truth). The view opposite to emergentism called \g{spiritualism} hypothesizes that, quite to the opposite, everything originates from people mind or from an omnipotent and omniscient being. Laplace was reported to have said on that respect: \fr{\g{L'hypothèse d'un être omnipotent explique tout en effet, mais ne permet pas de prédire quoi que ce soit}.}

Anaximander lived about six hundred years before our era. He was as well as Thales at the origin of the scientific way of thinking and the first to write what could be called a scientific treatise. This treatise, as well as all pre-socratic contributions, were unfortunately lost. The prevalent opinion at his time was that things have a natural tendency to \g{fall}, a way of thinking that begs the question. Why is not the earth falling itself? Anaximander answer was that the earth is isolated in space and therefore does not have by itself any tendency to move; but it attracts other objects. Why this is so is a question that was not answered then, and remains unanswered. We adopt the model of a flat earth, or a cylinder of very large diameter (according to modern concepts the diameter of this cylinder should be on the order of many light-years, so that a falling corpuscle could reach almost the speed of light; however, in usual circumstances, corpuscle speeds are considerably smaller than the speed of light). This amounts to saying that in our model the earth gravity may be considered a universal constant, independent of space and time: If a corpuscle is freed from some altitude $z_m$ its motion $z(t)-z_m$ does not depend on $z_m$.

\paragraph{Digression concerning Anaximander assertion that the earth thickness is one third of its diameter:}

According to later commentators, Anaximander viewed earth as a body with the shape of a cylinder of height (or thickness) equal to one third of the diameter. We offer here an explanation of that factor of 3. We suppose that Anaximender knew (e.g., from the shadow of earth in moon eclipses) that earth has the shape of a sphere, and that Anaximender assertion was that in his model the flat earth had a thickness equal to one third of the real-earth diameter. If this is so, it follows indeed from the Newtonian laws that the flat-earth model and the real spherical earth provide the same force of gravity on the surface, supposing the matter density to be the same (for details, see the french version of this paper).

\paragraph{Democritus concepts:}

Democritus distinguishes two forms of knowledge: knowledge from our senses, which he calls \g{obscure}, and knowledge from reason, which he calls \g{truthful}. As far as physics is concerned his view point is as follows: The universe consists of elementary corpuscles moving in vacuum, regularly or irregularly as a result of their mutual encounters. Democritus considered that inalterable and point-like corpuscles were moving in an infinite vacuum. They could not be directly observed, but they should be able to explain observables. This view point was well understood by Aristotle who wrote: \g{Democritus omits to discuss final causes, and therefore reduces to necessity all of the nature actions}. It seems that Democritus did not ascribe a weight to corpuscles. Indeed, one may think of a gravity model consisting of a \g{rain} of extremely high speed tiny particles moving in all directions and colliding elastically with the corpuscles. The concept that corpuscles have a weight of their own seems to have been introduced thereafter by Epicurius. The Democritian concept of corpuscles moving in vacuum was accepted by a few ancient philosophers and engineers such as Hero of Alexandria who, incidentally, invented the first thermal engine: the \g{eolipile}, a sphere containing water with oblique outlets heated by fire, which can be called a steam engine.

Epicurius and the latin poet Lucretius adopted the democritian view-point mainly for ethical reasons. Here are a few quotations attributed to Democritius on that respect:\\
\g{Freedom of expression is part of freedom}\\
\g{Laws should not forbid people to live as they please, as long as they do not harm others}\\
\g{Some, ignoring the dissolution of our nature, trouble their lives by fear of what may happen after their death}.

One can say with Salem \cite{salem} \fr{\g{Étrange destin que celui d'une physique qu'on ne saurait réduire à une anticipation chanceuse des dogmes sur lesquels reposent aujourd'hui notre science. Étrange destin que celui d'une éthique dont l'actualité paraît si évidente}.} Democritus teaching was generally ignored, rejected, or scorned until the $16^{th}$ century. It was written in 1670: \g{The corpuscular hypothesis, that is the opinion according to which the world is a fortuitous motion of atoms, is impious and awful}. But probably Democritus could have said with Régine Desforges (Le Cahier Volé) those words that reconcile in us feelings of unity and necessity: \fr{\g{J'aimerais mourir à l'automne et que mon corps enfoui à même la terre humide et encore chaude de l'été se décompose rapidement, participant ainsi à l'énorme travail de pourrissement qui accompagne tout renouveau.}} 

The Democritus concept interested many modern thinkers such as Diderot, Nietzsche, Marx, but also Leibnitz. The corpuscular hypothesis has often been considered as a mechanism that does not relate to philosophy as an ontology, namely the study of beings. Diogene Laerce ($\sim$ 200 of our era) observes that Plato feigns to ignore Democritus \cite{nikseresht}. The \fr{\g{parti pris des choses}} mentionned earlier defines a \emph{method} and an empathy rather than an ontology.

Most scientists today concur with the democritian hypothesis and employ instead the words: philosophy and metaphysics in a derogatory manner as  relating to language and to an historical tradition but lacking true significance. From the Democritian view point, those problems nonetheless persist. How things appear beyond the mechanisms of perception? and conciousness beyond neural phenomena? Where the notion of \g{meaning} is coming from? What do we call \g{imagination} and \g{love}? And as far as ethics is concerned, it seems insufficient to relate it to a simple mechanism of social animals evolution. These are problems related to emergence. Let us cite Kim \cite{kim} \fr{\g{Au fur et à mesure que les systèmes acquièrent des degrés de plus en plus élevés de complexité organisationnelle, ils présentent de nouvelles propriétés qui, en un certain sens, transcendent les propriétés de leurs parties constitutives et dont l’existence ne peut être prédite à partir des lois gouvernant les systèmes plus simples.}}

The interpretation of quantum mechanics require concepts having a philosophical connotation such as realism, locality, causality. The most common notion of causality is poorly understood by most in spite of its obvious usefulness, because many phenomena have multiple causes that are difficult to separate out. It is often not appreciated that strong correlations between sets of events needs not imply that one is the cause of the other. Besides, the widely-held belief in a direct action of our desires on matter and the belief in supra-natural actions confuses our understanding of things (\emph{de rerum natura}). Reasonings then proceed only by analogy, or superficial similarities. Does the Democritus theory possess a true scientific meaning and does it anticipate the modern atomic science? We believe that this is so, but others deny that this is the case. 

Let us recall that works on the air properties occurred in two periods at the modern time. The first begun with Torricelli, inventor of the barometer, led to the barometric experiments by Pascal (whose name incidentally is now used as the unit for pressure) in 1648. R. Boyle was then the first to publish in 1662 the experimental law according to which the product of pressure and volume of air is approximately a constant at room temperature. This law was later on found to be independent of the temperature and of the nature of the gas. We observe that the Boyle observation just cited may be implemented with nothing else but a glass tube (sealed at one end) and mercury. Glass and mercury were both available at the time of the ancient Greeks. If such experiments were performed at these early times, the results have not been transmitted to us. 

En 1738, D. Bernoulli (perhaps pre-dated in the modern time by Gassendi) gave to this law an atomic explanation: \fr{\g{les atomes d'un gaz se déplacent de façon aléatoire et la pression n'est rien d'autre que l'impact des atomes sur les murs du récipient qui contient ce gaz.}} It is interesting to observe that the word \g{gas} comes from the greek word meaning: chaos. Dalton has shown that chemical reactions occur in proportional parts. In particular, Avogadro gave the formula H+O+H for water. This conclusion rests on the idea that there exist a number (about 100) kinds of atoms called \g{elements}, atoms of the same kind being identical if one ignores isotopes. Boltzmann \cite{cercignani} was a major actor in introducing the atomic concept in thermodynamics. What is moving us most concerning the pre-socratic thinkers, more than their scientific achievements, is their \emph{heroism}. The books by Zeller and Salem, in particular, express that empathy. The reader is invited to look at the simulations available on the web\cite{simulations}.\\

Besides the introduction and the conclusion this work consists of five chapters, namely:\\
Chapter \ref{math} recalls the elementary mathematical operations required, particularly those applicable to integers: compositions.\\
Chapter \ref{pro} introduces the subject of thermodynamics through an hypothetical discussion between Democritus and the king of his country.\\
Chapter \ref{eqther} discusses thermal equilibrium between two constant-volume objects.\\
Chapter \ref{res} presents a particular kind of heat engine consisting of two reservoirs of corpuscles.\\
Chapter \ref{parf} treats in a general manner ideal gases, on the basis of the round-trip time of a corpuscle thrown upward in the earth gravity. We consider a single corpuscle whose energy evolves slowly because of its encounter with the hot ground level. The barometric law is also discussed and generalized to altitude-dependent weights.\\

Various functions and variables are employed in thermodynamics text-books. In Chapter \ref{pro} we evaluate the work and efficiency of an engine without using explicitly the concept of entropy. The most usefull function in Chapter \ref{parf} is the free-energy $A(\theta,h)$ (sometimes denoted $F(\theta,h)$). Here the temperature $\theta$ is introduced on dimensional grounds but is proved later on to be an absolute temperature. $h$ (or $\eps$) are parameters that may be varied almost at will when the working medium is in contact with a heat bath or is displaced from one bath to another. Given the free-energy function $A$, we can readily obtain the internal energy $U$, the average force $\ave{F}$ (or pressure), and the entropy $S$. In alternative formulations, $U$ is expressed as a function of entropy and a parameter. Then $\theta$ is the derivative of $U$ with respect to $S$ for a constant value of the parameter.


\chapter{Mathematics}\label{math}

\emph{After explaining the notation and terminology we list results that we shall need in this work relating to integers, for example the associativity of addition, and recall the definition of the compositions of a number. More advanced concepts such as that of the derivative of a function are also recalled.}

\section{History}\label{history}

More than 4 000 years ago Sumerians possessed advanced knowledge in mathematics: position numeration with a base 60 (instead of the bases 10 and 2  employed today), solution of second-degree equations, for example. This heritage was partly transmitted to the ancient Greeks, who augmented our knowledge in geometry. As an example, the formula giving the volume of a cone (one third of the product of the base area and the height) was suggested by Democritus and proven by Eudoxus.
 
Unfortunately, some operations, although possible in principle for the ancient Greeks, were in fact impracticable because the position numeration was not generally known. The use of \g{0} in calculations appeared in India \emph{circa} 500 of our era. Besides, the concept of cartesian coordinates or more advanced ways of representing space, see \cite{chappell} (where further historical accounts may be found) were not known to them. Combinatory analysis plays a fairly important role in this work. The fact that the number 103 049, announced by Hipparcus (\emph{circa} 150 before our era) in a rather obscure context found an interpretation in modern times \cite{combinatoire}, suggests that this astronom had non-trivial knowledge in combinatory analysis. 

\section{Notation and terminology}\label{not}

Concerning numbers, we use the usual position numeration with a base 10, and the anglo-saxon convention of using a point after an integer part. We also separate large numbers in sections of three numbers by a blank for easier reading. Let us recall some terminology, originating from the greek language: \g{iso} means \g{same}; \g{chore} means \g{volume}; \g{bar} means \g{pressure}; \g{therm} means \g{temperature}; \g{adiabatic} means \g{does not go through the wall} (implied: the heat, if the wall is motionless), \g{entropy} means \g{transformation}. We will show that in the context of the present work adiabatic transformations are isentropic. Some mathematical rules and definitions are given below to help readers. For a large part of the present paper it suffices to know the elementary rules of arithmetics. Elsewhere, we employ a modern language. Parentheses are employed to denote the order in which operations must be performed, beginning with the inner ones. For example, $(a+b)+c$ means that one must first add $a$ and $b$, and then add $c$ to the result. Parentheses are employed also with an entirely different meaning to indicate the argument of a function. For example, in: $f(x)$, or $h(x)\equiv f(g(x))$, the parentheses mean that, given $x$, there exists a rule $g$ giving some number $g(x)$. Knowing that number, another rule, $f$, delivers the number denoted $f(g(x))$, or more concisely $h(x)$. 
Finally, parentheses are also employed to denote sequences. Integers are denoted (...-2,-1,0,1,2,...). We employ mainly positive integers (1,2,...) and non-negative integers (0,1,2,...) denoted: $\mathbb{N}$. Note that \g{...} means that the previous (often implicit) rule is to be continued. For example: 0,1,2... refers to all the non-negative integers. The signs employed are:
\begin{align}\label{signs}
&R\Longrightarrow S \textrm{  means that the statement } R   \textrm{  implies the statement   } S \nonumber\\
&a=b\qquad a~ \textrm{coincides with}~ b \nonumber\\ 
&a\ne b\qquad a~ \textrm{does not coincide with}~ b \nonumber\\
&a\propto b\qquad a~ \textrm{proportional to}~ b \nonumber\\
&a<b \mathrm{\ \ (or\ } b>a)\qquad a~ \textrm{is smaller than}~ b \nonumber\\
&a\le b \mathrm{\ \ (or\ } b\ge a)\qquad a~ \textrm{is smaller than or equal to}~ b \nonumber\\
&a\gg b\qquad a~ \textrm{is much larger than}~ b \nonumber\\
&a\ll b\qquad a~ \textrm{is much smaller than}~ b \nonumber\\
&a\equiv b\qquad \textrm{the expression }a\textrm{ coincides with the expression }b \nonumber\\
&a\approx b\qquad a~ \textrm{is approximately equal to}~ b \nonumber\\
&a\sim b\qquad a~ \textrm{is of the order of magnitude of}~ b \nonumber\\
&a+b\qquad \textrm{denotes the sum of}~a~\textrm{and} ~b \nonumber\\
&a\times b\qquad \textrm{denotes the product of}~a~\textrm{and} ~b \nonumber\\
&a!=1\times 2\times3\times...a,\quad \textrm{$a$ positive integer}; \quad 0!=1;\qquad 1/a!=0, \quad\textrm{$a$ negative integer}\nonumber\\
&A=\{a, b\}\qquad \textrm{means that the set $A$ exclusively contains}~a~\textrm{and} ~b \nonumber\\
&a\in A\qquad a~ \textrm{is an element of the set}~ A \nonumber\\
&a\notin A\qquad a~ \textrm{is not an element of the set}~ A \nonumber\\
&\infty \qquad \textrm{is an arbitrarily large number} \nonumber\\
& \textrm{min($a$,$b$) is the smallest of $a$ or $ b$}\nonumber\\
&C(a,b)\equiv\frac{a!}{b!(a-b)!} \textrm{   equals 0 if}~ a<b.
\end{align}

\section{Elementary rules.}\label{matha}\label{elementary}

We use letters such as: $p,q...$ to represent non-negative integers whose values are not yet known. By construction:
\begin{align}\label{assio1}
(m+n)+p=m+(n+q)\qquad m+n=n+m,
\end{align}
so that the parentheses may be omitted. For example (1+2)+3=1+(2+3)=6. A negative integer (-$q$) is formally defined by the rule: $q$+(-$q$)=0.

A product $m\times n$ means that $m$ must be added to itself $n$ times, or conversely that $n$ must be added to itself $m$-times. For example: $2\times 3=2+2+2=3\times 2=3+3=6$. We have: 
\begin{align}\label{prod}
m\times n=n\times m\equiv    m\,n      \qquad m\times (n+p)=m\times n+m\times p.
\end{align}
We call: $m\times m\equiv m^2$ the square of $m$, or more generally $m^q$ the $q^{th}$ power of $m$; that is $m$ multiplied by itself $q$ times. By convention $m^0=1$. 

A sum of terms is denoted as follows:
\begin{align}\label{sum}
\sum_{n=0}^{n=N} a_n\equiv a_0+a_1+...+a_N.
\end{align}
For example $\sum_{n=1}^{n=3} n\equiv 1+2+3=6$. It follows from a previous rule that, with $N≥h$:
\begin{align}\label{sumg}
\sum_{n=0}^{n=h-1} a_n+\sum_{n=h}^{n=N} a_n\equiv a_1+a_2+...+a_{h-1}+a_{h}+a_{h+1}+...+a_{N}=\sum_{n=0}^{n=N}a_n.
\end{align}
We denote by $m^n$ the product of $m$ by itself $n$-times. By convention $m^0=1$. It follows that 
\begin{align}\label{assio}
m^{(n+p)}=m^nm^p\equiv m^pm^n; \quad m^{(np)}=(m^n)^p; 
\quad m^qm^{-q}=m^{q-q}=1\Rightarrow m^{-q}=1/m^q.
\end{align}
For example, $3^{(2\times 2)}=3^4=81$ and $(3^2)^2=9^2=81$.

The ratio of $m$ and $n$ is a rational $p\equiv \frac{m}{n}\equiv m/n$ such that $n\,p=m$. The number $p$ exists, except when $n=0$ and $m$ is non zero. When both $n$ and $m$ are 0 the ratio is not defined. The basic algebraic rules are the same as those applicable to integers.

Any number (called \g{real}) may be approximated as closely as we wish by rationals. The rules \eqref{sum}, \eqref{sumg}, \eqref{assio} are also applicable to rationals. We denote $\sqrt{m}$ a non-negative number $n$ whose square is $m$, that is, if: $n^2=m$, then $n\equiv\sqrt{m}$. The ancient Greeks demonstrated that $\sqrt{2}$ is not rational. However, this non-rational may be approximated as closely as we wish by rational numbers, e.g., $17/12<\sqrt{2}< 18/12$. Hero of Alexandria has shown that $\sqrt{2}$ may be obtained by recurrence from: $2x(n)x(n+1)-x(n)^2=2$ (which is an identity if $n+1\approx n$), beginning for example with: $x(2)=17/12$.

We have: 
\begin{align}\label{app}
2^\Delta-1\approx 0.69\, \Delta.
\end{align}
 when $\Delta$ is small compared with unity, as one can see by setting $\Delta=\frac{1}{q},~q\gg 1$ and calculating that: $2\approx (\frac{0.69}{q}+1)^q$ or $2\,q^q\approx (q+0.69)^q$. For example with $q=10$ we obtain numerically: $ (10.69)^{10}\approx 1.97\,10^{10}$,  which differs only slightly from $2\,10^{10}$.
 
It is convenient at that point to introduce a constant $e\equiv 2.718...\approx 2^{1/0.69}$. We then have the simpler relation: 
\begin{align}\label{e}
e^\Delta-1\approx \Delta, \qquad \Delta\ll1.
\end{align}
From the above rules, going from $2^x$ to $\exp(x)\equiv e^x$ merely amounts to a rescaling of $x$.

\section{Compositions}\label{compositions}

A \emph{composition} of an integer $n$ (sometimes called a \g{weak composition}) is a sequence of non-negative integers (called the \emph{parts}) whose sum is $n$. The number of compositions of $n$ with $g$ parts is denoted: $\Om(g,n)$. For example for $g=2$, (0,2), (2,0), (1,1) are the three compositions of 2. We will further suppose that the parts are taken in some non-empty subset $A$ of the non-negative integers. (Physically the subset $\{0,1\}$ is applicable to single-spin electrons, the subset $\{0,1,2\}$ to electrons, and the subset $\{0,1,...\}$ to bosons).

The number of compositions of $n$ with $g$ parts taken in $A$ is denoted: $\Omega_A(g,n)$. For example, $\Omega_{\{0,1\}}(2,2)$=1. There is indeed only one composition of 2 in two parts taken in $\{0,1\}$, namely: (1,1). Clearly, $\Omega_A(1,n)=1$ if $n\in A$ and $\Omega_A(1,n)=0$ otherwise. In words, there is only one composition of $n$ with a single part, namely ($n$), if part $n$ is allowed, and no composition if $n$ is not allowed. The values of $\Omega_A(g,n)$ may then be obtained by recurrence:
\begin{align}\label{recurA}
\Omega_A(g,n)=\sum_{a\in A}\Omega_A(g-1,n-a). 
\end{align}
beginning with $g=2$. Indeed we may split the $g$ parts into a first part and $g-1$ other parts. If the first part is $a$ (taken in $A$) the number of compositions of the rest is: $\Omega_A(g-1,n-a)$. Then we must sum over all allowed $a$-values. Note that $\Omega_A(g,0)=1$ if $0\in A$, because there is only one way of obtaining 0 as a sum of $g$ non-negative integers, namely $g$ zeros, and that $\Omega_A(g,0)=0$ if $0\not\in A$. 
Besides, $\Omega_A(g,n)=0$ if $n<0$ because one cannot obtain a negative integer by adding non-negative integers.

For example the above relation \eqref{recurA} reads:
\begin{align}\label{recur}
\Omega_{\{0,1\}}(g,n)&=\Omega_{\{0,1\}}(g-1,n)+\Omega_{\{0,1\}}(g-1,n-1)\nonumber\\
\Omega_{\{0,1,2\}}(g,n)&=\Omega_{\{0,1,2\}}(g-1,n)+\Omega_{\{0,1,2\}}(g-1,n-1)+\Omega_{\{0,1,2\}}(g-1,n-2)\nonumber\\
\Omega_{\{0,1,...\}}(g,n)&=\Omega_{\{0,1,...\}}(g-1,n)+\Omega_{\{0,1,...\}}(g-1,n-1)+...+\Omega_{\{0,1,...\}}(g-1,0)\nonumber\\
&=\Omega_{\{0,1,...\}}(g-1,n)+\Omega_{\{0,1,...\}}(g,n-1),
\end{align}
with:
\begin{align}\label{reur}
\Omega_{\{0,1\}}(1,n)&=1\qquad n=0,1\qquad \textrm{and 0 otherwise}\nonumber\\
\Omega_{\{0,1,2\}}(1,n)&=1\qquad n=0,1,2\qquad \textrm{and 0 otherwise}\nonumber\\
\Omega_{\{0,1,...\}}(1,n)&=1. 
\end{align}

The expressions:
\begin{align}\label{reclur}
\Omega_{\{0,1\}}(g,n)&=\frac{g!}{n!(g-n)!}\nonumber\\
\Omega_{\{0,1,2\}}(g,n)&=\sum_{n/2≤s≤min(n,g)}\frac{g!}{(g-s)!(2s-n)!(n-s)!}\nonumber\\
\Omega_{\{0,1,...\}}(g,n)&=\frac{(g+n-1)!}{ n!(g-1)!} 
\end{align}
satisfy the above conditions. In particular, the first expression in \eqref{reclur} satisfies the first expressions in  \eqref{reur} and \eqref{recur}. Indeed:
\begin{align}\label{test}
\frac{1!}{0!(1-0)!}=\frac{1!}{1!(1-1)!}=1\qquad \frac{1!}{n!(1-n)!}=0\quad n\notin \{0,1\}
\end{align}
remembering that the factorial of 0 is 1, and that the reciprocal of the factorial of a negative integer is 0. Next, we verify the recurrence relation (first in \eqref{recur})
\begin{align}\label{tst}
\frac{g!}{n!(g-n)!}=\frac{(g-1)!}{n!(g-1-n)!}+\frac{(g-1)!}{(n-1)!(g-n)!}.
\end{align}
remembering that: $g!/(g-1)!=g$. Likewise, we readily verify the last expression in \eqref{recur}. 

We have also: $\Omega_{\{1,2,...\}}(g,n)=\frac{(n-1)!}{ (n-g)!(g-1)!}$, whose sum over $g=1,2..$ gives $2^{n-1}$, a rather obvious result (place $n$ "1" in row, and separate them by either plus signs or spaces). One may also find in the literature \cite{combinatoire} (main menu, combinatorics: "balls in bins with limited capacities") the number of ways of placing $n$ corpuscles (or \g{balls}) in $g$ bins.

We observe for later use that:
\begin{align}\label{ggenA}
\frac{\Omega_{\{0,1\}}(g,n+1)}{\Omega_{\{0,1\}}(g,n)}=\frac{g-n}{n+1}\approx \frac{g}{n}-1\equiv \frac{1}{\nu}-1\quad n\gg1\\
\frac{\Omega_{\{0,1,...\}}(g,n+1)}{\Omega_{\{0,1,...\}}(g,n)}=\frac{g+n}{n+1}\approx \frac{g}{n}+1\equiv \frac{1}{\nu}+1 \quad n\gg1.
\end{align}
Concerning $\Omega_{\{0,1,2\}}(g,n)$, we have for example: $\Omega_{\{0,1,2\}}(3,4)=6$ and $\Omega_{\{0,1,2\}}(4,4)=19$. We obtain numerically with $g=20 000, ~n=10 000$ (that is, $\nu\equiv n/g=0.5$): $\Omega_{\{0,1,2\}}(g,n+1)/\Omega_{\{0,1,2\}}(g,n)$=2.3025... 

An expression for the number of compositions of $n$ in $g$ parts in $A=\{0,1,...R\}$ is:
\begin{align}\label{ggenC}
\Omega_{\{0,1,...R\}}(g,n)=\sum_{0≤s≤min(g,n/(R+1))}(-1)^s~ C(g,s)~C(g+n-(R+1)s-1,g-1)
\end{align}
where $C(a,b)\equiv \frac{a!}{b!(a-b)!}$. An expression is known also when the $g$ sites have different capacities $R$.

In the sequel of the present section we keep $g$ fixed and do not show it as argument. We define the temperature reciprocal: $\beta=\ln\p\Om(n+1)/\Om(n)\q$ or equivalently: $\exp(\beta)=\Om(n+1)/\Om(n)$. Two media $A$ and $B$ have therefore almost the same temperature if: $\Om_A(n_a+1)/\Om_A(n_a)\approx\Om_B(n_b+1)/\Om_B(n_b)$. 

If the total number of corpuscles is $n=n_a+n_b$ and the two media may exchange corpuscles the quantity $\Omega_A(n_a) \Omega_B(n-n_a)$ peaks sharply around a particular value of $n_a$ corresponding approximately to the equality of the temperatures defined above. For example, if $A=\{0,1\}$ and $B=\{0,1,...\}$, $g=n=50$, the product peaks for $n_a=15$, $n_b=35$. The temperature reciprocals are $\beta_A\approx\beta_B\approx 0.8$.

\section{Advanced rules}\label{adv}

We define integration as the sum of small quantities, as was done by Archimedus to evaluate the area between a parabola and a straight line. In the limit where $a_n$ does not vary much from one $n$-value to the next, we write:
\begin{align}\label{intn}
\sum_{n=0}^{n=N} a_n\equiv a_0+a_1+...+a_N\approx \int_0^N dn~a(n).
\end{align}
However, we use rarely integrals in this work.

The concept of the derivative of a function was introduced by Leibnitz and Newton. For example, one says that the derivative of the function $f(x)=x^2$ is $2\,x$ because $\frac{(x+\eps)^2-x^2}{\eps}=\frac{(x^2+2x\,\eps+\eps^2)-x^2}{\eps}\approx 2\,x$ if $\eps\ll x$. It follows from \eqref{e} that:
\begin{align}\label{inth}
\frac{df(x)}{dx}=f(x)\qquad \textrm{if}\qquad f(x)=\exp(x)\approx 2.718^x
\end{align}
The above holds only if $f(x)=e^x$. 

For a power $n$ of $x$ we have:
\begin{align}\label{ibnth}
\frac{dx^n}{dx}=n\,x^{n-1}\qquad  \int_0^y \,dx\, x^n= \frac{y^{n+1}}{n+1}\qquad n≥0.
\end{align}

The $\ln(x)$ function is the inverse function of the function $\exp(x)$, meaning that: $\ln(\exp(x))=x$. It follows that $\frac{d\ln(y)}{dy}=\frac{1}{y}$, as one can see by setting $y=\exp(x)$, taking the derivative of that expression with respect to $x$ and using \eqref{inth}. Double derivatives are denoted: $\frac{d}{dx}(\frac{df(x)}{dx})\equiv\frac{d^2\,f(x)}{dx^2}$.

The sign $\frac{\partial f(x,y)}{\partial x}$ simply means that variables other than $x$ are held constant. For any continuous function $f(x,y)$ of $x$ and $y$ we have:
\begin{align}\label{ih}
\frac{\partial}{\partial x}\frac{\partial f(x,y)}{\partial y}=\frac{\partial}{\partial y}\frac{\partial f(x,y)}{\partial x}\equiv \frac{\partial^2\,f(x,y)}{\partial x\partial y}
\end{align}
This rule may be verified for example for $f(x,y)=x^2\,y$ with the result $\frac{\partial^2\,f(x,y)}{\partial x\partial y}=\frac{\partial^2\,f(x,y)}{\partial y\partial x}=2x$. A small variation of $f$ is:
\begin{align}\label{intbbh}
df=\frac{\partial f(x,y)}{\partial x}dx+\frac{\partial f(x,y)}{\partial y}dy.
\end{align}

If a quantity Q whose small variation $\delta Q=h(x,y) dx+g(x,y)dy$ with $\frac{\partial h(x,y)}{\partial y}\ne \frac{\partial g(x,y)}{
\partial x}$, then $Q$ is not a function of $x$ and $y$. If we consider a closed path in the $x,y$-plane, the sum of the increments $\delta Q$ needs not vanish. The result depends on the path. This is the reason why we employ the notation $\delta Q$ instead of $dQ$.

\chapter{Classical Thermodynamics}\label{pro}

\emph{By \g{classical thermodynamics} we mean a theory that does not employ microscopic concepts. We first present an hypothetical dialogue between Democritus and the king of his country, Abdere. The king idea is to lift heavy loads with the help of a bag full of air heated by wood combustion. The question he raises is the following: how much heat (or more concretely, how much wood) is required to perform that task? Here we establish the ideal-gas law but employ an empirical expression for the energy required to heat air to some temperature. We discuss the Otto cycle, and explain next the Carnot concept (1824) relating to the maximum efficiency of a heat engine. The concept of entropy is not employed}. 

\emph{At the time of Lavoisier, it was generally believed that heat is a weightless fluid capable of flowing from hot bodies to cold bodies. This weightless-fluid theory (we refrain from calling it a \g{caloric} theory to avoid a confusion with the Carnot work) is applicable to cavities that contain only light, an almost weightless fluid. This fluid can be transferred from a hot body to a cold body, exert forces on a piston and deliver work. However, the number of light particles is not a constant in a closed vessel (a similar situation occurs for water vapor in equilibrium with liquid water). The number of corpuscles is constant when they have a mass, except at extremely high temperatures where particle-antiparticle pair creation may occur. If we put aside these fluids one should say, following Plato, that heat is motion and nothing else, or better that heat is a form of energy. This is the view-point adopted by Carnot who wrote: \fr{\g{La chaleur n'est rien d'autre que la puissance motrice, ou plutôt une autre forme du mouvement. Lorsque de la puissance motrice est détruite, la chaleur est générée précisément en proportion de la quantité de puissance motrice détruite; de même, quand de la chaleur est détruite, de la puissance motrice est générée.}}}

\newcommand{\dia}{\vspace{1em}\noindent---\hspace{1em}}
\section{The king idea and Democritus response.}

\dia Dear Democritus, said the king, I am building a castel. To perform that task, my workers have to lift heavy stones. Even with small-slope ramps hundreds of men are required to bring the stones extracted from the mountain to the locations defined by my architect. Having observed the expansion of tight air-filled bags under the influence of heat, I wondered whether fire, obtained for example by burning wood, could help us lift these stones with little human effort. Some arrangements are probably better than others. Being aware of the depth of your thinking on various matters in physics and philosophy I am expecting from you proposals.

\dia What you are proposing, replied Democritus, appears to me to be feasible indeed. To give you a precise answer, we could of course make a large number of tests, with various bags, heat consumption, and loads. I will show however that pure reasoning enables us to predict a large number of useful facts, so that only few tests will be needed. I only need the following measurement: When we let a known weight drop into a well-insulated container of some known height, I need to know how much heat is generated. This heat is measured by the equivalent quantity of burnt wood, that is, the quantity of wood that would give the same temperature rise. I do not know sufficiently well the nature of air to predict the value of this quantity (related to the \g{air heat capacity}).

I know that you possess a cylinder made of bronze sheets, with vertical axis, usually employed to collect rain water. I am proposing to employ as a length unit the height of this cylinder, and to call it: one meter. I noted that the cross-section area of that cylinder happens to be 1 meter $\times$ 1 meter. The bottom of that cylinder rests on the ground. The cylinder being filled of air in the usual conditions, its top is a piston which could possibly move up and down. Some oil prevents air from leaking in or out. My fundamental philosophical concept is that this cylinder contains a fixed number ($N$) of independent corpuscles. This number is probably huge, but I will not need to know it. What I need to know is that the force exerted on the piston is proportional to $N$, and that, of course, if a single cylinder contains $N$ corpuscles, two identical cylinders contain $2N$ corpuscles.

On the other hand, we have made the following experiment: We filled up with water a cylinder similar to the one just described, but with a height of 11 meters, and closed the upper end, letting the water escape from the lower end into a bassin. A water height of 10 meters remained in the cylinder. I conjecture that the weight of the remaining water is balanced by the weight of the atmosphere. I am then proposing to employ as a weight unit the weight of that remaining water\footnote{In modern units this weight is 10 tons or 100 000 newtons. The international system of mechanical units consists of the meter (length), kilogram (mass) and second (time). The unit of force (Newton) is the force required to communicate a speed of one meter per second in one second to a unit mass initially at rest. The energy is the product of one newton and one meter. Other energy units are the kilowatt-hour, the calorie, and, from this work view-point, the Boltzmann constant. Note that the words \g{energy} and \g{free-energy} are employed by some people with obscure meanings, unrelated to those given to these names in physics}. This atmospheric weight is a force permanently applied to the cylinder upper end. We may augment this force by adding weights on the piston, or reduce it by pulling up the piston.

At last, let us convene to call \g{energy} the product of a weight and the height by which it is raised. Simple experiments involving cords and well-oiled pulleys show that at equilibrium the total energy is a constant. We will see that when heat is involved, energy also remains constant, but it is not so easy to define \g{heat energy}.

\dia Of course, I have some understanding of what people mean by \g{temperature}. But I presume that you can give a more precise definition of that word.

\dia Having observed that, as the sensed temperature increases the total load that the cylinder piston may support without moving increases, I propose to call \g{absolute temperature} $T$ the weight of that load, plus the already-present atmospheric force. For example, the melting ice absolute temperature is 1 in those units and the boiling water absolute temperature is $T=1.37$. We could define similar temperature scales by employing other substances but air, for example a piece of copper; but these other scales may differ from the one just defined. This is for reasons having to do with the maximum efficiency of heat engines, reasons that will be later explained, that the temperature defined from air as described above may be called an: \g{absolute temperature}. (Note: This assertion would be more accurate if helium rather than air were employed. However the gas helium was not known to the ancient Greeks, even though air contains about 1\% of helium).

\dia Fine! said the king. So much being granted, what will you be able to demonstrate?

\dia On the basis of the corpuscular theory and some plausible assumptions I can establish the law of gases as I shall explain in a moment. As I said earlier, as far as the energy required to heat air to some temperature I need an empirical result. 

\paragraph{The ideal gas law:}

Let us consider again a cylinder with vertical axis in standard conditions: $h=F=T=N=1$, where $h$ is the cylinder height, $F$ the force exerted on the piston, and $T$ the temperature, and the number $N$ of corpuscles is set to 1 for mathematical convenience (any other fixed number would do). Let us multiply the temperature by $k$. At equilibrium the total load $F$ is multiplied by $k$ according to our definition of the temperature, $h$ and $N$ remaining the same. We set: $F=N\,f(T,h)$ and are looking for the function $f(T,h)$ of the two variables $T$ and $h$, with: $f(T,1)=T$. 

Let us now put one above another $h$ identical cylinders so that the total height is the integer $h$, with a load $k$ placed on the upper cylinder piston. The load is of course transmitted to the lower cylinders (we neglect the piston and gas weights), so that each cylinder is now submitted to a load $k$ and its height remains $h=1$, so that the temperature $T=k$. We assume that removing intermediate pistons is un-consequential (this is plausible according to our corpuscular model because there is a huge number of corpuscles and the corpuscles do not attract each other). If we consider the cylinder assembly as a whole, its temperature $T=k$, its height is $h$, the force is $k$, and $N=h$. Thus the general form $F=N\,f(T,h)$ reads: $k=h\,f(k,h)$ for any integral value of $h$, a relation that shows that $f(k,h)=k/h=T/h$. Finally, the relation between the quantities considered above is:  
\begin{align}\label{llo}
F=N\,f(T,h)=N\,\frac{T}{h}.
\end{align}

(Note: This \g{ideal-gas law} has been discovered on empirical grounds by Boyle who established that, for air and standard temperature, pressure$\times$volume is a constant. It is not difficult experimentally to show that the same relation holds, at least approximately, for any gas and at any temperature, e.g., any boiling-liquid temperature. This empirical result in conjunction with the theoretical result in \eqref{llo} shows that a vessel filled at standard pressure and temperature with any gas contains the same number of corpuscles. It is conventional to consider vessel volumes of 0.0224 cubic meters and call the amount of gas in it a \g{mole}. $N$ is then called the Avogadro number $N_A$.)

To establish the above law we have not used any microscopic considerations since the numerical value of $N$ remains arbitrary. But two assumptions were crucial: One is that the number of corpuscles $N$ in a closed vessel is a constant. In particular the above result is \emph{not} applicable to light, because the number of light particles is not a constant in a closed vessel. In that case $F$ does not depend on $h$, thus violating \eqref{llo}. Second, we have assumed that removing intermediate pistons is un-consequential, an assumption that would not hold for few corpuscles and interacting corpuscles. The above argument to obtain the ideal-gas law is thus not circular (or tautological) since, if it were, the above conditions would be superfluous while, as we have shown, they are crucial.

\paragraph{How to obtain the maximum efficiency:}
We will employ below the more general form:
\begin{align}\label{lo}
F=T g(h),
\end{align}
where $g(.)$ is not necessarily proportional to $N/h$, but we set $g(1)=1$ so that $T$ remains defined as before. As you will see, this formula suffices to establish an expression of the maximum efficiency of thermal engines.

Thanks to the river flowing near Abdere, we have at our disposal in winter time melting ice, whose temperature, from the adopted definition, is unity. On the other hand, by burning wood we may heat gases to a temperature that I denote $T$. At that temperature $T$ the force is $F(h)=T\,g(h)$ according to \eqref{lo}. If the cylinder height $h$ is incremented by $\Delta h$, the elementary amount of work (or energy) is: $F\,\Delta h=T\,g(h) \Delta h$. Thus, if the height is incremented from $h_1$ à $h_2$, the temperature being maintained constant, the total work performed is: $W_h=T\,\Sigma(h_1,h_2)$, where $\Sigma(h_1,h_2)\equiv \Sigma$ is a sum of elementary terms, whose expression will not be needed. Let now the cylinder be cooled down to the temperature $T=1$ without changing $h$, then brought back to its initial height (thus from $h_2$ to $h_1$). The work supplied is now $W_l=\Sigma(h_2,h_1)=-\Sigma(h_1,h_2)=-\Sigma$. The $\Sigma$ are opposite to the previous ones since the same work elements are taken in the opposite direction. The net work performed is therefore: $W=W_h+W_l=(T-1)\Sigma$. (Note: $\Sigma$ is the integral of $g(h)$ from $h_1$ to $h_2$. For ideal gases this quantity is $\ln(h_2/h_1)$ which may assume arbitrarily large values).

The amount of heat delivered by the hot bath is equal to $W_h$ plus the amount of heat required to raise the gas temperature from 1 to the temperature $T$ for some fixed value of $h$. We assume that this quantity does not depend on $h$ and thus does not grow to infinity when $h$ increases. The term $W_h$ can be made as great as we desire by increasing the gas expansion $h_2/h_1$, while the amount of heat is bounded. It follows that in that limit the amount of heat required to heat the gas is negligible, and the efficiency of the machine, defined as the ratio of the work delivered over a full cycle to the heat supplied by the hot bath, may be written:
\begin{align}\label{prolo}
\eta_C\approx  \frac{W_h+W_l}{W_h}=\frac{T-1}{T}=1-\frac{1}{T}.
\end{align}
This expression gives the value of the maximum (Carnot) efficiency. In the case of boiling water: $T=1.37$, and the maximum efficiency is thus: $\eta_C=1-1/1.37\approx 0.26$. A unity efficiency is reached only if the cold bath absolute temperature is 0.

\paragraph{Work produced by ideal gases:} In order to obtain the performance of the system you are envisioning, Your Majesty, I need describe more accurately the system operation. I will show that for ideal gases, see \eqref{llo}, the optimum load is about 1/5 (that is 2 tons). Let me suppose that there is, besides the unity atmospheric force, a load $w$ on the piston, which is prevented to get lower than 1 meter. The cylinder is put in contact with the hot bath at temperature $T=1.37$. The piston altitude then increases from 1 up to some value $h$ given by $(1+w)h=T$ according to the ideal gas law \eqref{llo}. The load will be lifted provided $T>1+w$. The load is then displaced laterally and secured. The useful work performed $W$ is the product of $w$ and the height increment: $h-1$ (Note: we do not consider raising the atmosphere to be an useful work; this is why the work is $(h-1)w$, not $(h-1)(1+w)$). We have therefore: 
\begin{align}\label{prol}
W=w(h-1)=w(\frac{T}{1+w}-1). 
\end{align}
Setting the derivative of the above expression with respect to $w$ equal to 0, we find that $W$ is maximum when: $w=\sqrt T-1=\sqrt {1.37}-1\approx0.17$. I accordingly suggest that you lift by the procedure just described stones of weight close to 0.17 (that is 1.7 tons). The useful work performed will then be: $W=w\,(h-1)=(\sqrt{T}-1)^2\approx0.027$ in our units.

To conclude, by going from melting ice to boiling water temperatures, the machine that you are proposing will be able to raise weights of 0.17 (in our units) by 0.18 meters. This operation may be repeated as frequently as you wish. However one must proceed slowly in order for the contacted bodies temperatures to equalize.

\paragraph*{The smallest simple-machine heat consumption.}

Your goal however, Your Majesty, may not be to reach the maximum work in a single step, but rather to minimize the wood consumption for a total given work. Most of the time the load may be split into a number of parts, and the lifting operation may be repeated. I define the \g{consumption} as the ratio of the heating energy and the work performed. This the reciprocal of what I called earlier efficiency.
\begin{figure}
\centering
\includegraphics[width=0.8\columnwidth]{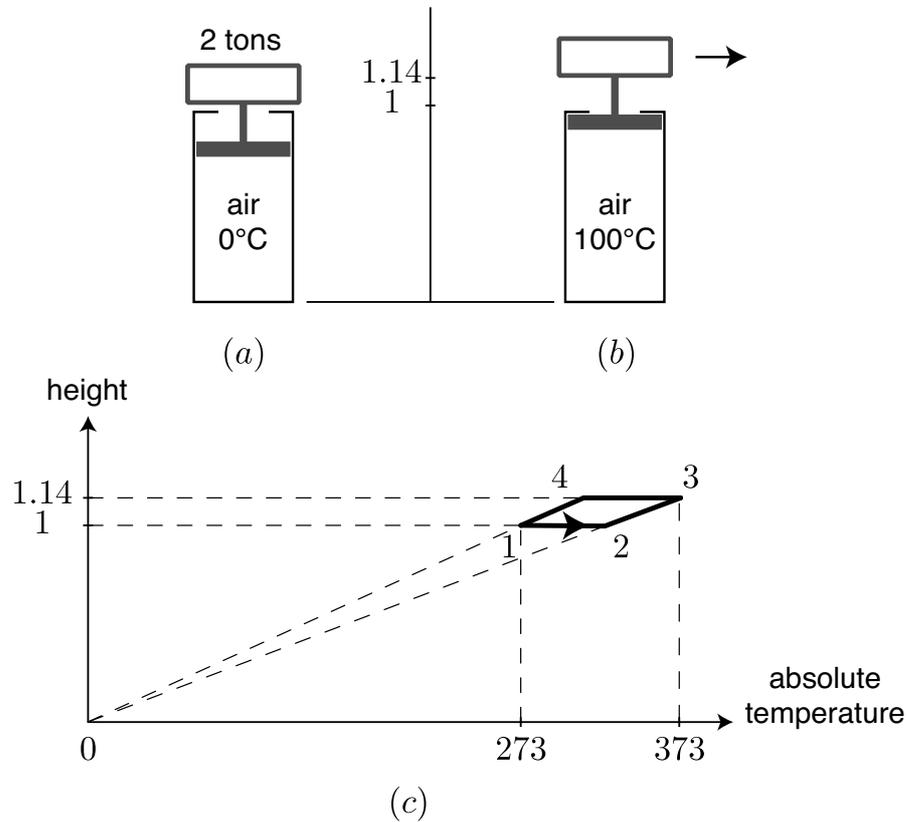}
\caption{This figure describes a simple machine consisting of a cylinder of unity cross-section area. The piston height with respect to the cylinder bottom is denoted $h$. In (a), we have represented the cylinder with a load of 2 tons. The temperature is that of melting ice (0 °C; read: degree celsius) and $h=1$ meter. In (b), the temperature is that of boiling water (100 °C) and $h$ became 1.14 meter. The load is displaced laterally. The piston is prevented from moving to a higher altitude. The diagram in (c) represents the variation of $h$ as a function of the air absolute temperature. This diagram involves two isochores (constant heights: 1-2 and 3-4) and two isobars (constant forces: 2-3 and 4-1). For graphical reasons we have supposed the cylinder axis vertical. This, however, is unimportant as long as $h$ is much smaller than about 1 km on earth.}
\label{roi}
\end{figure}
As far as the heating energy is concerned, I observe that heating one cubic meter of air (fixed volume) in standard conditions from temperatures 1 to $T$ may be obtained by burning wood. But it may also be obtained by dropping a weight in the cylinder from top to bottom (1 meter). The weight then defines the energy equivalent to some wood burning. Such experiments provide the so-called \g{internal energy}, $U(T)$, of one cubic meter of air, initially at standard pressure, at some temperature  $T$. The result is: 
\begin{align}\label{sf}
U=a\,T=2.3\, T
\end{align}
The modern values are the following: The mass of one cubic meter of dry air in the standard conditions is 1.2 kilogrammes. The amount of heat required to raise the temperature of one kilogram of air by one kelvin (later defined) for a constant volume is 710 joules. The melting-ice temperature is 273 kelvins. It follows that: $a=710\times1.2\times 273/100 000\approx 2.3$ in our units, because one atmosphere corresponds to 100 000 newtons for a unit area\footnote{When $U$ is proportional to $T$ as supposed here, $a$ is called the constant-volume heat capacity. Its theoretical value for one degree of freedom is 1/2, see later in Section \ref{internal}. For di-atomic nitrogen and di-atomic oxygen, the theoretical value of $a$ is (3+2)/2=2.5. It is customary to set: $\gamma \equiv 1+1/a$. This is the ratio of the constant-pressure heat capacity and constant-volume heat capacity. This constant is most easily measured from the speed of sound.}.

The thermal energy that one must supply is the work $(1+w)(h-1)$ that includes the work made against the atmospheric pressure, plus the increment $a(T-1)$ required to raise the air temperature from 1 to $T$. The useful work is, as said earlier, $W=w(h-1)$. The consumption is therefore, with: $W=w(h-1),~ (1+w)h=T$:
\begin{align}\label{sdf}
C(w) =\frac{(1+w)(h-1)+a(T-1)}{W}.
\end{align}
Substituting in that formula $T$ by 1.37 and $a$ by 2.3, we observe that the minimum consumption is: $C\approx36$. It is obtained for a value of $w$ slightly larger than the one providing the maximum work. The diagram height-temperature is shown in Fig.~\ref{roi}, consisting of two isochores (1-2 and 3-4) and two isobars (2-3 and 4-1).

I conclude than the consumption of the simple machine described (C=36) is 10 times larger than the smallest possible consumption relating to the same high and low bath temperatures. Unfortunately the optimum (Carnot) machine requires more involved mechanisms.

\section{The Otto cycle}\label{otto}

A simple machine may be implemented in a somewhat different manner. In that new cycle, the working medium (cylinder height $h$) is not modified when the cylinder is in contact with either baths. This cycle, as the previous one, is not reversible and therefore non-optimal \cite{arnaud1, arnaud2, arnaud3, arnaud4}. 

To obtain the work delivered during adiabatic steps we need both the ideal-gas law \eqref{llo} and the expression of the internal energy \eqref{sf}: $U=a\,T,~F\,h=T$. Since no heat may come in or out of the cylinder we must have: $0=dU+F\,dh=a\,dT+\frac{T}{h}dh$, or: $a\frac{dT}{T}+\frac{dh}{h}=0$. This implies that: $T^a\,h=$ constant, by taking the logarithm of that expression and the derivative. Replacing $T$ by $F\,h$, we obtain also: $F(h)\,h^{1+1/a} = \text{constant}$.

Suppose that initially $T=1$ (cold bath) and $h=1$; then $F=1$. Once the cylinder has been separated from the cold bath, $h$ is reduced from 1 to $h'$. The work performed is the integral of $F(h) \,dh$ from $h=1$ to $h'$. Thus, using: $F(h)\,h^{1+1/a}=1$, we obtain by integration:
\begin{align}\label{sjdf}
W_1=\int_1^{h'} \frac{dh}{h^{1+1/a}}=a\p 1-h'^{-1/a}\q.
\end{align}
The gas reaches a temperature $T'$ given by: $T'^a\,h'=1$, or: $T'=h'^{-1/a}$. According to \eqref{sf} the heat supplied by the hot bath is: $Q=a(T-T')=a(T-h'^{-1/a})$. It remains to evaluate the work performed during the second adiabatic step from $T,h'$ to $T'',1$ ($T''$ will not be needed). Initially $F=T/h'$. Thus $F(h)\,h^{1+1/a}=(T/h')\,h'^{1+1/a}=T\,\,h'^{1/a}$. $W_2$ is the integral from $h'$ to 1 of $F(h)dh$ with $F(h)$ just given. We obtain: $W_2=-W_1\,T\,h'^{1/a}$. The work performed and the efficiency are therefore, respectively:
\begin{align}\label{spf}
W&=W_1+W_2=a\p1- h'^{-1/a}\q\p1-T\,h'^{1/a}\q=a\p1- 1/x\q\p1-T\,x\q\nonumber\\
\eta&=\frac{W}{a(T-h'^{-1/a})}=1-x\qquad \qquad x\equiv h'^{1/a}\equiv  h'^{\gamma-1}.
\end{align}
If we vary $h'$ we find from the above formulas with that the maximum work occurs for $x=1/\sqrt{T}$. Then $\eta=1-1/\sqrt{T}<1-1/T,\,T>1$. Thus the efficiency at maximum work is less than the maximum (Carnot) efficiency.

\section{The Carnot theory}

Sadi Carnot raised \emph{circa} 1824 the following question: \cite{schroeder} What physical arrangement would make the best use of the available heat? His answer, based on the law of energy conservation, is that the best engine must be reversible, in the sense that, operating in the reverse manner, the same work would subtract the same amount of heat (heat pump). Indeed, if an engine could have a greater efficiency that a reversible engine, it would be possible by combining a heat engine and a heat pump, to generate more work than used initially, in violation of the law of energy conservation. Next, he noticed that the only fundamental source of irreversibility is the contact beween bodies at different temperatures. It is therefore necessary to avoid such contacts. Note that if the temperature difference between contacted bodies is vanishingly small, the rate at which heat is being transfered is also vanishingly small. It must therefore be realized that the \emph{power} generated by a reversible engine is vanishingly small. Carnot considered only a slow, or quasi-static, operation. The important quantities then are the work produced per cycle and the heat consumption. The basic laws are the following:

\begin{description}

\item[\No 0] This is the law of thermal equilibrium discussed in more detail later in Chapter \ref{eqther}.\\

\item[\No 1] This is the law of conservation of energy, including thermal energy. Carnot \cite{carnot} wrote in notes published after his early death:\g{La chaleur n'est rien d'autre que la puissance motrice, ou plutôt une autre forme du mouvement. Lorsque de la puissance motrice est détruite, la chaleur est générée précisément en proportion de la quantité de puissance motrice détruite; de même, quand de la chaleur est détruite, de la puissance motrice est générée}. He established (on partly empirical grounds) that the amount of heat required to raise one gram of water by 1 celsius is equivalent to 3.26 joules (instead of the modern value of 4.18 joules).\\

\item[\No 2] No work may be obtained through a cyclic operation from a constant-temperature bath. This law is sometimes expressed by saying that the \emph{entropy} of an isolated system never decreases. This law is thus based on the concept of entropy called \g{calorique} by Carnot. Let us quote on that respect Zemansky and Dittman \cite{zemansky}: \g{Carnot used chaleur when referring to heat in general, but when referring to the motive power of fire that is brought about when heat enters an engine at high temperature and leaves at low temperature, he uses the expression: chute de calorique, never: chute de chaleur. Carnot had in the back of his mind the concept of entropy, for which he reserved the term: calorique. Carnot had acquired the concept of entropy for which he reserved the name of calorique}. For that reason we can only approve the proposal made by a number of authors to give to the unit of entropy the name: Carnot (Cn in short) instead of the currently used: joule per kelvin. For a very complete discussion of the Carnot contribution, see Brodiansky \cite{brodiansky}.

\end{description}

By considering intermediate thermal engines Carnot has shown that the efficiency of reversible engines depends only on the hot and cold bath temperatures and is of the form:
\begin{align}\label{ffnj}
\eta_C\equiv \frac{\textrm{work performed}}{\textrm{heat subtracted from the hot bath}}=\frac{W}{Q}=1-\frac{\theta(T''_l)}{\theta(T''_h)}, 
\end{align}
where $\theta(.)$ is some unknown function of the temperature $T''$, which is defined on an arbitrary scale (e.g. through the expansion of some piece of metal). Later on we will only use the absolute temperature $\theta$. Clearly, $\theta$ is defined only to within an arbitrary constant factor.

\begin{figure}
\centering
\includegraphics[width=0.5\columnwidth]{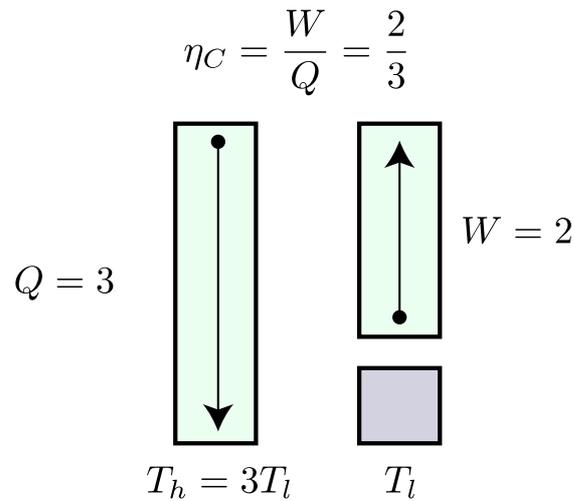}
\caption{This figure illustrates the result of the operation of an optimal (Carnot) heat engine. In this model the heat consumption $Q$ originates from a mechanical work; the fall of a weight $Q$. The representation on the left is schematic. It may represent a cylinder terminated by a piston and filled up with air. The diagram pressure-volume (or force-height) is in Fig.~\ref{carnotdecarnot}. The energy $Q-W$ is dissipated in the cold bath, which warms up. The processes are supposed to be arbitrarily slow. We also suppose that the baths are so large that their temperatures do not vary much cycle after cycle.}
\label{cyclecarnot}
\end{figure}

We give a simple illustration of the \emph{result} of a Carnot cycle on \ref{cyclecarnot}. The cold bath is supposed to have an absolute temperature equal to 100 and the hot bath an absolute temperature equal to 300. Carnot theory predict, according to \eqref{ffnj} an efficiency: $\eta=\eta_C=200/300$. The heat energy $-Q_h\equiv Q$ could be supplied by the fall of a weight over a height of 3 meters in the hot bath. Ideally, the machine could raise the same weight over 2 meters. Of course, this operation that converts a work into a lesser work does not present any practical interest. It is presented here to concretize the significance of the amount of heat $Q$. The difference of temperature between the ocean surface and deep waters may be on the order of 20 celsius (or kelvin). An optimum thermal engine operating these two baths, invented by G. Claude \cite{claude}, is quite low, on the order of one per cent. Yet, such an engine may prove useful.
 
\begin{figure}
\centering
\includegraphics[width=0.6\columnwidth]{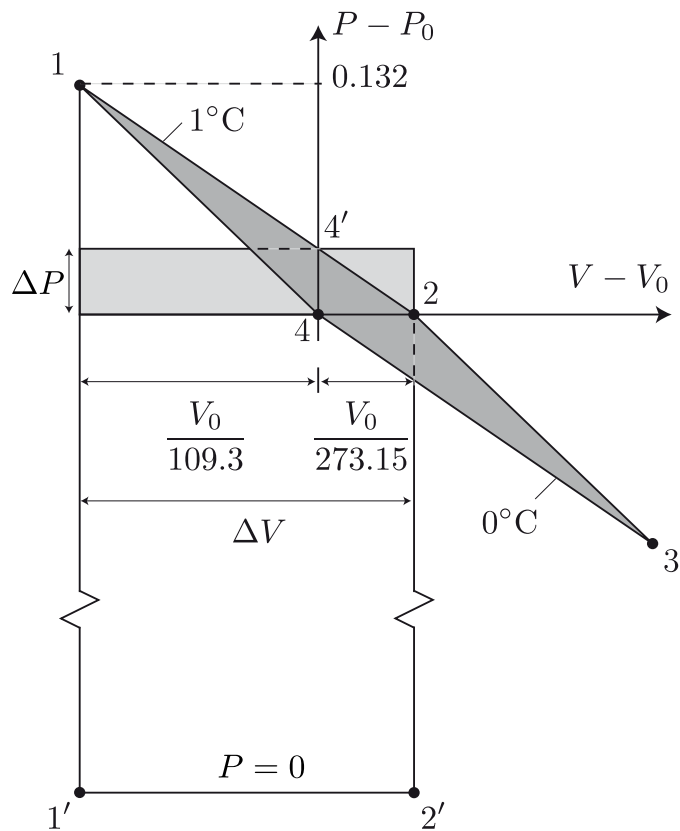}
\caption{This figure represents the Carnot cycle as described by Carnot (but not drawn exactly as we have done here) in his 1824 booklet. $P$ is the pressure and $V$ the volume. $P_o$ is a reference pressure, for example the atmospheric pressure, and $V_o$ a reference volume, for example one cubic meter. The cold-bath temperature is that of melting ice. The hot-bath temperature is one °C (or one kelvin) higher. The work delivered $W$ is the area of the cycle, performed in the direction: 1,2,3,4. This area is the same as that of the rectangular cycle shown on the left. However that rectangular cycle (isobar-isochore) would require a much larger amount of heat. The transformations 1-2 and 3-4 are \g{isotherm}: the temperature is maintained constant. The transformations 2-3 and 4-1 are \g{adiabatic}: heat does not go through the cylinder walls. They are also \g{isentropic}. Let us present the cycle in a more detailled manner. In 1 the gas is in contact with the hot bath. It is allowed to expand up to point 2 delivering work. From 2 to 3 the gas is forced to expend further, absorbing work, so that it cools down to reach the cold-bath temperature. Then from 3 to 4 it is allowed to compress, delivering work. Finally, the gas is forced to contract so that it reaches the hot bath temperature, absorbing work.}
\label{carnotdecarnot}
\end{figure}

\chapter{Thermal equilibrium}\label{eqther}

\emph{A common experience consists of contacting two bodies of constant volume having initially different temperatures (the two bodies may be two pieces of beryllium oxide that exchange energy through acoustical waves, or conductors such as pieces of copper that may exchange electrons). It is observed that the two temperatures eventually equalize, as one can judge by our senses, the final temperature being intermediate between the two initial temperatures. It is also found that the process is irreversible in the sense that if the pieces are separated, they do not recover their initial temperatures. In fact, they keep the same temperature. This irreversibility is the fundamental reason why a heat engine efficiency is less than unity, even if we neglect friction and other effects of the same nature. We will consider more particularly two media (a) and (b) having each $g$ distinguishable sites. In medium (a) each site may receive at most one corpuscle, that is 0 or 1, while in medium (b) the number of corpuscles per site is arbitrary: 0,1,2...The total number of sites is denoted $2\,g$ and the total number of corpuscles is denoted $2\,n$. We are studying the equilibrium between these two bodies with the help of a simulation. The numerical result may be obtained theoretically by evaluating the number of compositions of $2n$ with the first $g$ parts taken in the set \{0,1\} and the others in the set \{0,1,2...\}, and asserting that the number of such compositions is stationary when a corpuscle is transferred from (a) to (b), or the opposite. The equilibrium just described is called the law \No0 of thermodynamics. Under the conditions of slow evolution and in the absence of friction, thermal contacts between bodies having different temperature is the only source of irreversibility. In the present chapter the nature of the corpuscles, and in particular their weight, is irrelevant.}

\section{Simulation for two identical media}\label{i}

Following Ehrenfest  \cite{ehrenfest} one can give a plausible model of thermal contacts by using the corpuscular concept. This model has been illustrated with two dogs, say (a) and (b). One of the dogs carries $N$ fleas (representing corpuscles) while the other dog does not carry any. We select a flea at random and order it to jump on the other dog. This order is repeated a large number of times (in the first step a flea on dog (a) is of course selected. But later on there are fleas on both dogs). Under these conditions, we observe after a certain time the two dogs carry approximately the same number of fleas. Supposing that the temperature is a monotonically increasing function of the number of fleas, this model shows that the temperatures tend to equalize, in agreement with observations.

To be more accurate, we should observe that, even after a long time, the number of fleas on a dog will rarely be exactly equal to $N/2$, because fluctuations about this value perpetuate. Secondly, it may be that after a very long time, all the fleas are carried by dog (a) again. However, it has been shown that the number of steps for this to occur is on the average $2^N$. For large $N$-values, this time is so large that it can be considered infinite. The irreversibility mentioned before applies only to large-size objects.

From a physical viewpoint, note that the two dogs are supposed to be at the same altitude. Because we neglect the effort made by the fleas to jump from one dog to the other, the energy is constant. If the flea weight is unity and the altitude is unity, then the energy of a flea is unity, and the total energy equals the number of fleas.

\section{Simulation for two different objects}\label{diff}

Let us now consider two different objects, each involving $g$ sites. The sites are distinguishable and may be labeled from 1 to 100 for the first object and from 101 to 200 for the second object if $g=100$, for example. We suppose that the sites in object (a) may contain at most one corpuscle, that is, 0 or 1, while for object (b) the sites may accomodate any number of corpuscles, that is, 0,1,2...To obtain the condition of thermal equilibrium between these two objects, we perform the following simulation. We suppose that initially each site contains a single corpuscle, so that in that example the total number $2n$ of corpuscles equals $2g$. We select at random (with uniform probability) a corpuscle, and transfer it to a randomly-selected site that can accomodate this corpuscle, that is, an empty site for (a) and any site for (b). This procedure is repeated a large number of times, until a fairly stable situation is reached. For the purpose of selecting a corpuscle at random, the corpuscles may be labeled by their initial site number, from 1 to 200, and we require that the computer selects a number from 1 to 200. On a modern desk computer this procedure is very quick.

We obtain in that way an average number of corpuscles per site in (a): $\nu_a\equiv n_a/100=$ 0.383, and an average number of corpuscles per site in (b): $\nu_b\equiv n_b/100=$ 1.617. The sum of these two numbers is of course equal to 2 since the total number of corpuscles $n=n_a+n_b=200$ is unchanged. But we also have with good accuracy:
\begin{align}\label{ffn}
\frac{1}{\nu_a}-1=\frac{1}{\nu_b}+1, 
\end{align}
which suggests that the two members of this equality are related to the temperatures. A curiosity is that, since $\nu_a+\nu_b=2$,  the exact value of $\nu_b$ is the golden number: $(1+\sqrt5)/2$.

This simulation indicates what is the nature of a thermal equilibrium, but it does not allow us to define precisely a temperature. We will see later on that 
the two members of \eqref{ffn} are equal to: $\exp(1/\theta)$ where $\theta$ is the temperature (defined only to within a constant multiplicative factor). According to this formula, we have: $\nu_a=\nu_b=0$ if $\theta=0$, and $\nu_a=1/2$, $\nu_b=\infty$ if $\theta=\infty$. If $\nu_a$ were greater that 1/2 the temperature would be negative, but this cannot happen in the present case because the occupation of the sites does not have an upper bound. The concept of negative temperature will not be considered further in this work. 

If we have $n$ corpuscles in $g$ sites the number of distinguishable configurations is the number of compositions of $n$ with $g$ parts taken in some sub-ensemble $A$ of the integers (exhibited as a subscript). For two objects, the number of distinguishable configurations is the product of the numbers of configurations of the two systems: $\Om=\Om_A\,\Om_B$ since, to each composition of the first object we may associate a composition of the second object. At equilibrium, the transfert of a corpuscle from (a) to (b) should not affect significantly $\Om$. It follows that we must have:
\begin{align}\label{o}
\Omega_{A}(g_a,n_a)\Omega_{B}(g_b,n_b)&\approx \Omega_{A}(g_a,n_a-1)\Omega_{B}(g_b,n_b+1)\nonumber\\
\frac{\Omega_{A}(g_a,n_a)}{\Omega_{A}(g_a,n_a-1)}&\approx\frac{\Omega_{B}(g_b,n_b+1)}{\Omega_{B}(g_b,n_b)}\qquad n\gg 1.
\end{align}
For the objects considered we have (see Section \ref{math})
\begin{align}\label{om}
\Omega_A(g_a,n_a)\equiv\Omega_{\{0,1\}}(g_a,n_a)&=\frac{g_a!}{n_a!(g_a-n_a)!}\nonumber\\
\Omega_B(g_b,n_b)\equiv\Omega_{\{0,1,2...\}}(g_b,n_b)&=\frac{(n_b+g_b-1)!}{ n_b!(g_b-1)!}.
\end{align} 
Therefore:
\begin{align}\label{omv}
\frac{\Omega_{A}(g_a,n_a)}{\Omega_{A}(g_a,n_a-1)}&=\frac{g_a-n_a+1}{n_a} & \frac{\Omega_{B}(g_b,n_b+1)}{\Omega_{B}(g_b,n_b)}=\frac{g_b +n_b}{n_b+1}
\end{align}
Neglecting the \g{1} compared with $n_a$, $n_b$ we obtain from \eqref{o}
\begin{align}\label{ov}
\frac{g_a}{n_a}-1=\frac{g_b}{n_b}+1 && \Longrightarrow && \frac{1}{\nu_a}-1=\frac{1}{\nu_b}+1, &&  \nu_a\equiv \frac{n_a}{g_a},~\nu_b \equiv \frac{n_b}{g_b},
\end{align}
which coincides with \eqref{ffn}. The $\nu_{a,b}$ are called the average weights per site, supposing that each corpuscle has unit weight.

We have defined above the condition of thermal equilibrium between two objects, but we have not defined precisely the temperatures $\theta_a$ and $\theta_b$. To do so, it is useful to introduce an entropy function: $S\equiv \ln(\Om)$, so that the products in \eqref{o} are converted into sums. Here, $\ln$ denotes a natural logarithm (or logarithm of base $e$). Using any other logarithm, such as the logarithm in base $2$, would only amount to multiplying the temperature by some constant factor. This is unimportant since temperatures are defined only so far to within constant factors. Furthermore, we will only need entropy differences, so that an arbitrary constant may be added to $S$. It is usual to set $S=0$ at $\theta=0$ (this convention, justified in quantum theory, is usually called the third law of thermodynamics). With these definitions, the entropy is additive, that is, the entropy of two objects (in equilibrium or not) considered together, is the sum of their individual entropies: $S=S_a+S_b$.

We now introduce the concept of energy. If the reservoir of corpuscles, each of unit weight, is located at an altitude $\eps$ the corpuscle energy is $\eps$. From now on, we take $\eps=1$. We define the temperature reciprocal: $\beta\equiv 1/\theta=\Delta S$, the change of $S=\ln(\Om)$ when one corpuscle is added. We will show in the next chapter that $\theta$ so defined is an absolute temperature because the Carnot principle is obeyed, irrespectively of the nature of the objects.

\section[Average weight per site\dots]{Average weight per site deduced from the Boltzmann factor}\label{b}

In the previous section we considered a transfer of corpuscles from one site to another and employed a mathematical expression for the number of compositions of an integer with $g$ parts. We see here that a simpler solution may be obtained if we postulate the so-called Boltzmann factor. It is convenient to now suppose that the sites are located at an altitude $\eps$, a corpuscle at the ground level having an energy equal to 0 by convention. The corpuscles have a unity weight so that the energy of a corpuscle at altitude $\eps$ has an energy $\eps$. This energy is called \g{potential energy} in the sense that it could be converted into other forms of energy. 

Let us thus postulate that the probability for a corpuscle to have an energy $\eps$ is proportional to: $x\equiv\exp(-\eps/\theta)\equiv \exp(-\beta \eps)$. This is the \g{Boltzmann} factor. For simplicity we set: $\eps=1$. Then: $x\equiv\exp(-1/\theta)$.

\paragraph{A=\{0,1\}}

\begin{figure}
\centering
\includegraphics[width=0.6\columnwidth]{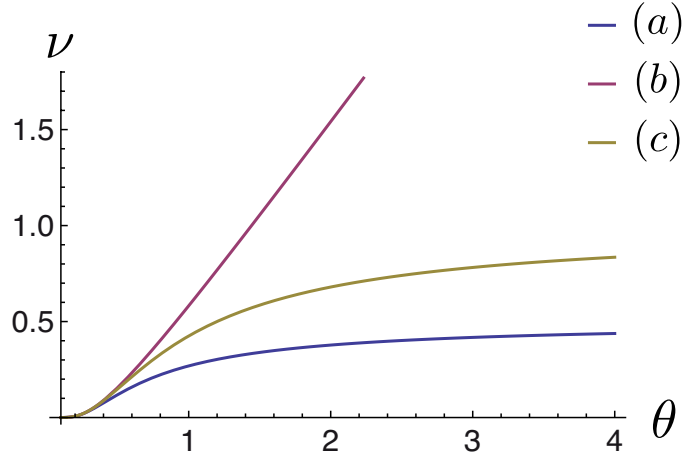}
\caption{Average energy per site $\nu_{a,b,c}$ as a function of the absolute temperature $\theta$. Medium (a): one corpuscle at most per site; (b): any number of corpuscles per site; and (c): two corpuscles at most par site. We suppose that the corpuscles have unit weight and are located at altitude $\eps=1$.}
\label{poids}
\end{figure}

In the case of medium (a) where the occupations of a site are either 0 or 1, the probability to have 0 or 1 are respectively proportional to 1 and $x$. Normalization to 1 gives:
\begin{align}\label{fk}
\nu_a=\frac{x}{1+x}\Longrightarrow \frac{1}{\nu_a}-1=\frac{1}{x}=\exp(1/\theta). 
\end{align}
The average occupation of a site (or average weight since the corpuscle weights are unity) is shown as a function of the temperature $\theta$ in Fig.~\ref{poids} in (a).

If we restaure $\epsilon$ we may write the above relation as:
\begin{align}\label{fkk}
\beta \eps =\ln\left(\frac{1}{\nu}-1\right). 
\end{align}
\paragraph{A=\{0,1,2...\}}

In the case of medium (b) (arbitrary occupation), the average weight per site is, according to the Boltzmann factor: 
\begin{align}\label{fhzk}
\nu_b=\frac{\sum_0^\infty n\,x^{n}}{\sum_0^\infty x^{n}}=\frac{x}{1-x}\Longrightarrow \frac{1}{\nu_b}+1=\frac{1}{x}=\exp(1/\theta).
\end{align}
For the first step, note that the geometric series $\sum_0^\infty x^{n}=1/(1-x)$, and take the derivative with respect to $x$. $\nu_b$ is shown as a fonction of temperature in Fig.~\ref{poids} in (b).

\paragraph{A=\{0,1,2\}}

Finally, consider a medium (c)  where each site may be occupied by two corpuscles at most. The probabilities to have 0, 1 and 2 corpuscles are proportional to $1,x,x^2$ respectively. Thus:
\begin{align}\label{kkmx}
\nu_c=\frac{x+2x^2}{1+x+x^2}. 
\end{align}
$\nu_c$ as given by the above expression is shown as a function of temperature in Fig.~\ref{poids} in (c).

For the medium (c) presently considered, the number of compositions with $g$ parts is, according to Chapter \ref{math}: 
\begin{align}\label{ccv}
\Omega_{\{0,1,2\}}(g,n)=\sum_{n/2≤s≤\min(n,g)}\frac{g!}{(g-s)!(2s-n)!(n-s)!}.  
\end{align}
We obtain numerically from this formula for $g$=20 000 and $\nu=n/g=1/2$: $1/x=\Omega_{\{0,1,2\}}(g,n+1)/\Omega_{\{0,1,2\}}(g,n)$=2.3025..., which agrees with the result obtained from the Boltzmann factor, see \eqref{kkmx}: $\nu=(x+2x^2)/(1+x+x^2), ~1/x=2.3...$.



\chapter{Reservoirs of corpuscles}\label{res}

\emph{The machine considered in this chapter consists of two reservoirs of corpuscles located at different altitudes, both with $g$ sites. If the corpuscle weight is nearly the same in the higher and lower reservoirs, the machine efficiency is nearly equal to the Carnot efficiency. It follows that the temperatures introduced are absolute temperatures. Only the concept of potential energy and average weight per site are employed.}

\emph{In order to enable the reader to comprehend this chapter without referring to previous ones, let us recall that the source of energy of a thermal engine is the heat $-Q_h\equiv Q$ delivered by a bath at some constant high temperature $\theta_h$. Part of it is converted into work $W$, the rest $Q_l=Q-W$ being dissipated in a cold bath at a constant low temperature $\theta_l$. The amount of energy supplied by the hot bath is the work that one should supply to maintain its temperature strictly constant, for example through the fall of a body into the viscous fluid. The transfer of heat between the hot bath and the cold bath is effected by a working substance (involving a varying parameter) alternately in contact with the hot and cold bathes. In the present chapter we call that parameter $\epsilon$ and identify it with the altitude of the site \No1. We consider particularly the Otto cycle where the parameter does not vary when the working medium is in contact with the heat baths. The machine that we are going to describe, made of two reservoirs containing identical corpuscles, operates in a manner similar to an Otto cycle, which is in general irreversible; but the Carnot efficiency $\eta$ may be reached in the limit of a large number of Otto cycles.}

\section{Two reservoirs of corpuscles}\label{deux}

The machine model presently considered consists of two reservoirs of corpuscles of unity weight located respectively at altitudes $\eps_l$ et $\eps_h>\eps_l$  in the earth gravity. Each reservoir contains $g$ identified sites labeled from 1 to $g$. The first site constitutes the working medium while the other sites describe the cold and hot baths, respectively. 

\begin{figure}
\centering
\includegraphics[width=0.3\columnwidth]{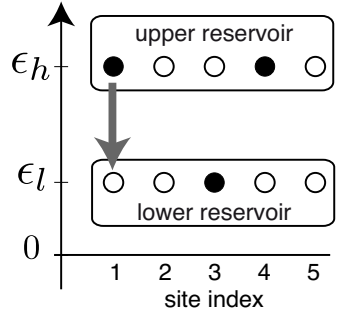}
\caption{This figure represents a model of thermal engine made of two corpuscle reservoirs located respectively at altitudes $\eps_l$ et $\eps_h>\eps_l$  in the earth gravity. Each site may be occupied by 0 or 1 corpuscle. A cycle consists of moving the site 1 from the upper to the lower reservoir, and back, allowing enough time in each step for an equilibrium situation to be reached. The site 1 may be considered as the working medium, while the other sites constitute the hot bath (higher reservoir) and cold bath (lower reservoir). The average work produced is positive in the case of the figure because there are more corpuscles (black circles) in the upper reservoir than in the lower one.}
\label{deuxreservoirs}
\end{figure}

The lower reservoir contains $n_l$ corpuscles and the upper reservoir contains $n_h$ corpuscles, all of unity weight. We set: $\nu_l\equiv \frac{n_l}{g}$, $\nu_h\equiv \frac{n_h}{g}$. These $\nu_l,\,\nu_h$ are called: average weights per site. A cycle consists of displacing the site 1 (working medium) from the lower reservoir (cold bath) to the upper reservoir (hot bath) and conversely, once an equilibrium situation has been reached.

\section{Work and efficiency}\label{mod}

If a corpuscle is added to the lower reservoir its energy is incremented by $\eps_l$. If a corpuscle is added to the upper reservoir its energy is incremented by $\eps_h$. Thus, after the exchange mentioned earlier, the average energies added to the reservoir are respectively: 
\begin{align}\label{add}
Q_l&=\eps_l(\nu_h-\nu_l)\nonumber\\
Q\equiv -Q_h&=\eps_h(\nu_h-\nu_l).
\end{align}
Indeed, the lower reservoir site receives on the average $\nu_h$ corpuscles, but it looses on the average $\nu_l$ corpuscles. For the higher reservoir the situation is opposite.

The average work performed by this arrangement is given by the law of conservation of energy: $Q_l+Q_h+W=0$, or: $W=Q-Q_l$ . It follows that the work produced and the efficiency are given by the following relations:
\begin{align}\label{donc}
W&=Q-Q_l=(\eps_h-\eps_l)(\nu_h-\nu_l)\nonumber\\
\eta&\equiv \frac{W}{Q}=1-\frac{\eps_l}{\eps_h}.
\end{align}
The significance of these formulas is illustrated in Fig.~\ref{cycle} that shows that the work, that is the difference between $Q$ and $Q_l$, is the area of the rectangle shown (for $\eps_{h2}=\eps_{h1},~\eps_{l2}=\eps_{l1})$. To be more concrete we could ascribe an electrical charge to the corpuscles and suppose that they are submitted to an electrical field instead of gravity. The energy exchanges are then made with the field source, e.g., a condensator that may get charged or discharged.

The above formulas seem to have little to do with temperature and fire. The relation existing between the apparently purely mechanical considerations given above and thermal effects is clarified below.

\section{Introduction of temperature }\label{int}

Let us define the entropy of a reservoir as equal to the number of sites $g$ multiplied by some function $s(\nu)$: $S(n)=g\,s(\nu)$, where as previously $\nu$ represents the average weight of a site. The fact that $S$ is proportional to $g$ implies that this quantity is additive: if $g$ and $n$ are both multiplied by 2 for example (thus letting $\nu$ unchanged) $S$ is multiplied by 2 as well.

Consider a reservoir located at the altitude $\eps$ and containing $n=\nu g$ corpuscles. Let us add one corpuscle. The temperature $\theta$ is defined as the energy increment divided by the entropy increment. The energy increment is $\eps$ since the weight of a corpuscle is unity while the entropy is incremented by: $\Delta S=S(n+1)-S(n)=g\,\p s(\frac{n+1}{g})-s(\frac{n}{g})  \q    \approx \frac{ds}{d\nu}$ if $n$ et $g$ are large. Indeed, $s(\frac{n+1}{g})\approx s(\nu)+\frac{ds}{d\nu}\frac{1}{g}.$ Thus: 
\begin{align}\label{t}
\theta&\equiv\frac{\eps}{\Delta S}=\frac{\eps}{ds/d\nu}\equiv\frac{\eps}{s'(\nu)}\nonumber\\
\beta \eps&=s'(\nu)\equiv f(\nu).
\end{align}
We see that $\theta$ does not depend on $g$: This is an intensive quantity. 

The expressions \eqref{donc} for the work and efficiency may now be written as follows
\begin{align}\label{doncbi}
W&=-\theta_l\,\Delta S_l-\theta_h\,\Delta S_h, &  \Delta S_l&=(\nu_h-\nu_l)f(\nu_l), & \Delta S_h=(\nu_l-\nu_h) f(\nu_h)    \nonumber\\
&&\eta&=1-\frac{\theta_l\,f(\nu_l)}{\theta_h\,f(\nu_h)}.&
\end{align}
We have introduced above the entropy increments of the two baths: $\Delta S_l$, $ \Delta S_h$ when site \No1 is being exchanged between the two reservoirs. 

\begin{figure}
\centering
\includegraphics[width=0.7\columnwidth]{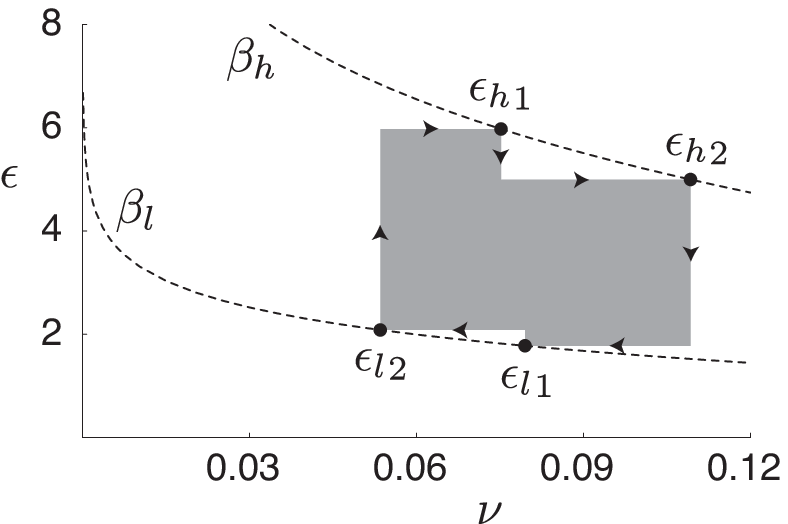}
\caption{ Given the low and high bath temperatures ($\beta_l=1.38;\quad\beta_h=0.42$), we draw the two curves $\eps_l=\theta_l f(\nu)$ and $\eps_h=\theta_h f(\nu)$ as functions of $\nu$. The figure represents a cycle involving two higher reservoirs and two lower reservoirs. The rectangular cycle considered in the text (Otto cycle) corresponds to the special case: $\eps_{h2}\to\eps_{h1},~\eps_{l1}\to\eps_{l2}$. The more complicated cycle shown here suggests the way Carnot cycles may be described, considering essentially a very large number of adjacent Otto cycles. The work delivered is the cycle area shown in grey.}
\label{cycle}
\end{figure}

In the special case where: $\nu_l\approx\nu_h$, we have: $f(\nu_l)\approx f(\nu_h)$, and thus: $\Delta S_l+\Delta S_h\approx 0$: the total entropy generated is almost zero. We set: $\Delta S\equiv \Delta S_l\approx -\Delta S_h$: the machine is almost reversible. The expressions of work and efficiency become:
\begin{align}\label{doncbis}
W&=(\theta_h-\theta_l)\Delta S\nonumber\\
\eta&=1-\frac{\theta_l}{\theta_h}.
\end{align} 
These are the formulas given by Carnot. They do not depend on the entropy function selected. In the limit considered, the work performed vanishes. However, we may always consider a sequence of rectangles each with $\nu_l\approx \nu_h$, such that we go through the picture in Figure \ref{cycle} from the Otto cycle to the Carnot cycle with some work being produced. The fact that the temperature $\theta$ as defined above leads to the Carnot efficiency: $\eta_C=1-\frac{\theta_l}{\theta_h}$ suffices to prove that $\theta$, as defined above, is an absolute temperature.

To summarize, let us rewrite the fundamental relations \eqref{add}:
\begin{align}\label{ad}
Q_l&=\theta_l f(\nu_l)(\nu_h-\nu_l)\nonumber\\
Q&=\theta_h f(\nu_h)(\nu_h-\nu_l).
\end{align}
For a heat engine we select $\nu_h>\nu_l$. For a heat pump, we select $\nu_h<\nu_l$, corresponding to a negative work. From the above relations we obtain the work performed (or absorbed) and the efficiency (or its inverse, the coefficient of performance) provided the function $f(\nu)\equiv s'(\nu)$ be known.

We consider below the case where the site \No1 (which serves as a working agent) may be occupied by one corpuscle at most, that is, 0 or 1.

\section{One corpuscle at most per site }\label{fermions}

In the more specific model presently considered we suppose that the two reservoirs (labeled respectively $l,\,h$) $g$ sites may be occupied at most by one corpuscle. If a reservoir contains $n$ corpuscles we set: $\nu\equiv \frac{n}{g}$. It is intuitive that the parameter $\nu$ represents the average weight of a site. We are restricting ourselves to positive temperatures: $\nu_l<1/2$, $\nu_h<1/2$. However, the two reservoirs are not in thermal equilibrium among themselves at some positive temperature since the upper one has a greater weight than the lower one: $\nu_h\ge\nu_l$, as shown in the figure. This of course is the same in conventional heat engines: the hot and cold baths are not in thermal equilibrium. This is precisely the reason why work may be extracted from them in a cyclic operation.

To obtain the numerical value of the work delivered according to the general form \eqref{doncbi}, we must know the values of $\nu_l$, $\nu_h$ and the function: $s'(\nu)\equiv ds(\nu)/d\nu$, the temperatures being supposed known. Let us recall (see the expression in \eqref{fk}) that if a site may be occupied by one corpuscle at most:
\begin{align}\label{fermibis}
\nu=\frac{1}{1+\exp(\beta \, \eps)}\qquad\beta \, \eps\equiv f(\nu)=\frac{ds(\nu)}{d\nu}=\ln \p \frac{1}{\nu}-1 \q\qquad\beta\equiv \frac{1}{\theta}.
\end{align}
To within an arbitrary additive constant the entropy is:
\begin{align}\label{entro}
S=g\,s(\nu),\qquad s(\nu)\equiv-(1-\nu)\ln(1-\nu)-\nu\ln(\nu). 
\end{align}
In our model constant values of $\nu$ correspond to isentropic transformations. Our representation differs however from the traditional $S,T$ (or $T,S$) representation in that $\eps$ is not simply proportional to $T$.

As an example, suppose that the reservoir altitudes are $\eps_l=1$ et $\eps_h=2$. If $\theta_l=1$ et $\theta_h=2$ we obtain from \eqref{fermibis} the values $\nu_l=\nu_h=1/(1+e)$ with $e\approx 2.718$. In that limit the efficiency is optimal and equal to 1/2. To obtain a non-zero work it is however necessary that $\nu_l$, $\nu_h$ be not strictly equal.

\begin{figure}
\centering
\includegraphics[width=0.5\columnwidth]{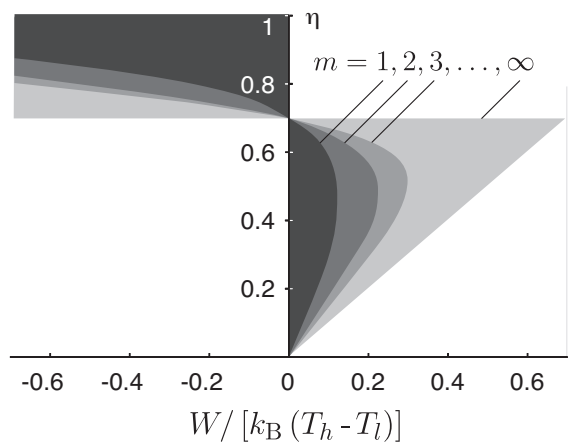}
\caption{ This figure shows the engine efficiency $\eta$ as a function of the work produced $W$ for one higher and one lower reservoirs (Otto cycle: $m=1$), two higher and two lower reservoirs ($m=2$),..., until the Carnot cycle is reached( $m=\infty$). Negative work corresponds to heat pumps. We have set: $\beta_l=1.38;\quad\beta_h=0.42$. Each couple of $\nu_l,\,\nu_h$ values give points: $\eta,\,W$ that cover the black area (for $m$=1). }
\label{eta}
\end{figure}

\chapter{Ideal gases}\label{parf}

\emph{We call \g{ideal gases} gases made up of corpuscles that do not interact with each other, but may collide with the walls of the cylinder in which they are enclosed, and then slowly gain or loose energy (this model is directly applicable to molecules in a planet such as Mercury having a rarefied atmosphere and a ground temperature on the order of the melting-ice temperature). The number $N$ of corpuscles in a closed vessel does not vary. This fits well with the Democritus model according to which everything is made of corpuscles moving in vacuum (see the introduction). This model suffices to explain the observation made by Boyle that the product of the average force $\ave{F}$ exerted by the gas on a piston terminating the cylinder multiplied by the height $h$ of the cylinder is a constant; this is so at any (various substances triple-points) temperature. In the present chapter treatment it is unnecessary to suppose that plates may be removed at will as we did earlier in Chapter~\ref{pro}. The product $\ave{F} h$ may be called an absolute temperature, denoted $\theta$. The generalized Boyle law holds true independently of the gas considered (say, helium or air). The Democritus model indicates that $\theta$ is proportional to the number $N$ of corpuscles contained in the cylinder. Accordingly, for simplicity, the theory presented below assumes that a single corpuscle is present.} 

\emph{Even so the corpuscle weights are ignored in the usual formulation of the ideal gas law, we find it convenient to consider corpuscles submitted to some gravity, e.g., the earth gravity, so that the \g{energy} of a corpuscle is most easily defined. Further, in some cases (kilometer-high cylinders) the role of the earth gravity is significant. The precise law of motion of a corpuscle submitted to gravity, however, is unnecessary. To the contrary, we employ a principle of simplicity asserting that the ideal gas law and the barometric law must be \emph{independent} of the laws of motion. For that reason, the principle reported in this work could be taught in a physics course \emph{before} any considerations concerning corpuscle motion are presented to students. The ideal gas law may fail to hold at very low temperatures at which quantum effects become relevant, or at very high temperatures when the corpuscles considered may dissociate or get ionized. The gas internal energy is derived.}

\section{Impact concept}\label{fh}

Let us first consider a unity-weight corpuscle moving along the vertical $z$-axis, and bouncing off the ground elastically on the plate of a balance, in a periodic manner. The average weight is unity irrespectively of the maximum altitude $z_m$ reached by the corpuscle, or, in other words, irrespectively of the corpuscle energy, denoted $E$. One can define the \g{impact} $i$ of the corpuscle on the balance as being equal to the motion period $\tau(E)$. Indeed, the average weight is equal to the impact $i$ multiplied by the number of impacts by unit time, thus to $i/\tau$. It follows that $i=\tau$ since we have selected a unit weight. The impact $i$ has the dimension of the product of a mass and a speed called \g{change in momentum}. In classical mechanics the impact is twice the momentum of the incident corpuscle. Here we do not suppose that the corpuscle momentum after a bounce is opposite to its incident momentum. Furthermore, we do not make the non-relativistic approximation.

\begin{figure}
\centering
\includegraphics[width=0.7\columnwidth]{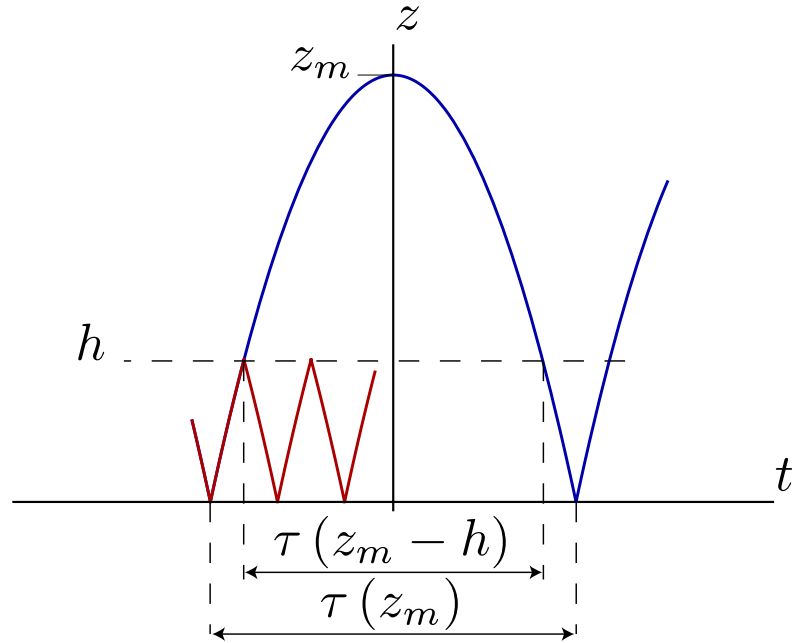}
\caption{ This figure represents the motion $z(t)$ of a corpuscle in the earth (or Mercury) gravity with maximum altitude $z_m$, according to the usual (Einstein) clock synchronization method. The dashed line represents a piston at altitude $h$. This figure refers to a zero temperature environment, in which case the corpuscle energy $z_m$ is a constant.}
\label{figure}
\end{figure}

\begin{figure}
\centering
\includegraphics[width=0.7\columnwidth]{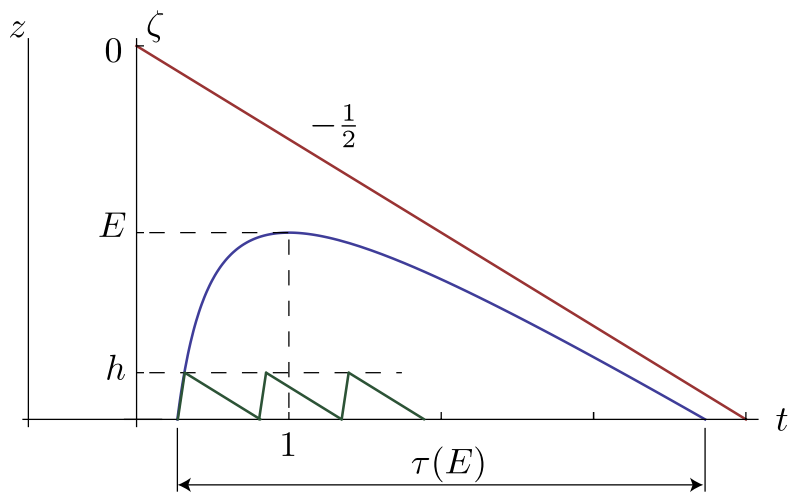}
\caption{ This figure represents the motion $z(t)$ of a corpuscle for an (unusual) clock synchronisation method: clocks are synchronized by a pulse emitted from the origin. It differs from the one shown in \ref{figure} only when the non-relativistic approximation is not made. The dashed line represents a piston at altitude $h$. $\zeta$ is a shifted $z$-axis. The line with slope -1/2 is an aid to construction. Our purpose is to show that, as far as the laws of thermodynamics are concerned, the method of clock synchronisation employed is immaterial, a non-obvious fact in usual presentations: Only corpuscle round-trip times matter.}
\label{parfait}
\end{figure}

\section{Force }\label{retour}

Let us now introduce a rigid plate at $z=h\ll E$. We have seen that the impacts are proportional to the time periods. These periods are obviously equal on the two planes. We will first avoid the use of integrals and employ only integers, supposing that the corpuscle energy, previously denoted $E$, is an integer $m$, so that the arithmetic results given in Chapter \ref{math} suffice. We will need consider at the end the large-temperature limit to remove the effect of energy discretization, which is made here for mathematical not physical reasons. As said earlier, Quantum theory calls for a uniform quantization of the \emph{action} (area in the $z,t$ space) not of the energy.

Consider first the case where $h$ is infinite, that is, in the absence of a piston. The average force exerted on the ground, equal to the corpuscle weight $w=1$, is the impulse ${i}$ divided by the period. Thus $1={i}/\tau(m)$ or ${i}=\tau(m)$. In other words, the impulse transmitted to a piston when the corpuscle impacts on it is equal to the motion period. If the piston is located at $h$ the impulse it receives is: ${i}_h=\tau(m-h)$ if $m\ge  h$, and zero otherwise. On the other hand, the motion period is $\tau(m)$ if $m\le h$ and $\tau(m)-\tau(m-h)$ if $m>h$. 

Accordingly, the force $F$ experienced by the piston when the corpuscle energy is $m$ is the ratio of the impulse and period:
\begin{equation}\label{orcebis}
\begin{cases}
F(m)=0 &m<h,\\
F(m)=\frac{i_h}{\tau(m)-\tau(m-h)}=\frac{\tau(m-h)}{\tau(m)-\tau(m-h)} \qquad &m\ge h.
\end{cases}
\end{equation}

Because the cylinder lower end is in contact with a bath at some non-zero temperature (e.g., the ground), there is a slight quivering (thermal motion) and the corpuscle energy $m$ slowly varies in the course of time. Accordingly, the average force $\ave{F}$ experienced by the piston is, if $\om(m)$ denote the energy distribution:
\begin{align}\label{mo}
\ave{F}=\frac{\sum_{m=0}^{m=\infty} \, \om(m)F(m)}{\sum_{m=0}^{m=\infty}\,  \om(m)}=\frac{\sum_{m=h}^{m=\infty} \, \om(m)\frac{\tau(m-h)}{\tau(m)-\tau(m-h)}}{\sum_{m=0}^{m=h-1} \, \om(m)+\sum_{m=h}^{m=\infty}\, \om(m)}.
\end{align}

According to our simplicity principle, the average force $\ave{F}$ must be \emph{independent} of the corpuscle equation of motion, and thus of the $\tau(.)$-function. This condition obtains from \eqref{mo} if one selects the following energy distribution:
\begin{equation}\label{pond}
\begin{cases}
\om(m)=\exp(-m)\tau(m) &m<h,\\
\om(m)=\exp(-m)\p\tau(m)-\tau(m-h)\q \qquad &m\ge h,
\end{cases}
\end{equation}
The average force becomes, using \eqref{mo} and \eqref{pond}:
\begin{align}\label{moy}
\ave{F}&=\frac{\sum_{m=h}^{m=\infty}\, \exp(-m)\tau(m-h)}{\sum_{m=0}^{m=h-1}\,  \exp(-m)\tau(m)+\sum_{m=h}^{m=\infty}\, \exp(-m)\p\tau(m)-\tau(m-h)\q}
\nonumber\\
&=\frac{\exp(-h)}{1-\exp(-h)}=\frac{1}{\exp(h)-1}
\end{align}
In the above sums going from $h$ to $\infty$ we have replaced $\exp(-m)$ by $\exp(-h)\exp(-(m-h))$ and introduced the variable $m-h$, so that all the sums go from zero to infinity and cancel out. We have employed also the fact that $f(x)=c^x$ is the only function such that $f(a+b)=f(a)f(b)$ for any constant $c$. 

For corpuscle weights $w$, the force is multiplied by $w$ and the energy becomes $w\,h$. We therefore replace $h$ in the above expression by $\beta\,w\,h\equiv w\,h/\theta$, where $\theta$ has the dimension of an energy since the argument of the exponential must be dimensionless. Thus, the above expression may be written:
\begin{align}\label{moyyg}
\ave{F}=\frac{w}{\exp(\beta\,w\,h)-1}=\frac{1}{\beta}\frac{\partial \ln(Z)}{ \partial h}\qquad Z(\beta,h)=\p \exp(-\beta\,w\,h) -1\q f(\beta),
\end{align}
where $f(.)$ is any function, introduced for reasons that will appear in the next section.

For a collection of $N$ independent corpuscles having weights $w_i,~i=1,...N$ respectively, the force is a sum of $N$  terms of the form given in \eqref{moyg}. In the case of zero gravity ($w$=0 or more precisely: $wh\ll \theta$), the above expression gives: $\ave{F}h=\theta$. Thus we have obtained the ideal-gas law: $\ave{F}h=N\,\theta$.

For a three-dimensional space, we suppose that the cylinder radius is very large compared with $h$, and we do not consider the force exerted by the corpuscle on the cylinder wall. Motion of the corpuscle along directions perpendicular to $z$ (say, $x$ and $y$) \emph{does} affect the round-trip time function $\tau(Z)$. However, since the average force does not depend on this function, the ideal-gas law is unaffected. This is so for any physical system involving a single corpuscle provided the  physical laws are invariant under a $z$-translation besides being static.

The internal energy, to be discussed in the following section, though, is incremented. One can prove that in the non-relativistic approximation and in the absence of gravity the internal energy is multiplied by 3. It would be incremented further by corpuscle rotation or vibration, not considered here. Using conventional methods, Landsberg \cite{landsberg} in Eq.\,(2.6), and Louis-Martinez \cite{Louis-Martinez} in Eq.\,(68), obtain the result given above (except for the factor 3 in the expression of the internal energy, relating to the number of space dimensions considered) in the continuum limit.

\section{Internal energy}\label{internal}

The gas internal energy $U$ is the average value of $E=m$, the average being calculated with the distribution in \eqref{pond}. Remember that only corpuscule motion along the $z$-axis is being considered. We proceed as in the previous section replacing $m$ by $m-n+n$ and thus $\exp(-m)$ by $\exp(-n)\exp(-(m-n))$. We obtain:

\begin{align}\label{auuu}
U&=\frac{\sum_{m=0}^{m=h-1}\, m\,\exp(-m)\,\tau(m)+\sum_{m=h}^{m=\infty}\,m\,\exp(-m)\p\tau(m)-\tau(m-h)\q}
{\sum_{m=0}^{m=h-1}\, \exp(-m)\,\tau(m)+\sum_{m=h}^{m=\infty}\,  \exp(-m)\p\tau(m)-\tau(m-h)\q}\nonumber\\
&=\frac{\sum_{m=0}^{m=\infty}\, m\,\exp(-m)\tau(m)-\exp(-h)\sum_{m=h}^{m=\infty}\, (m-h+h)\exp(-(m-h))\tau(m-h)}
{\sum_{m=0}^{m=\infty}\,  \exp(-m)\tau(m)-\exp(-h) \sum_{m=h}^{m=\infty}\, \exp(-(m-h))\tau(m-h)}\nonumber\\
&=\frac{(1-\exp(-h))\sum_{m=0}^{m=\infty}\,m\,\exp(-m)\tau(m)-h\,\exp(-h)\sum_{m=0}^{m=\infty}\, \exp(-m)\tau(m)}
{(1-\exp(-h))\sum_{m=0}^{m=\infty}\, \exp(-m)\tau(m)}\nonumber\\
&=\frac{\sum_{m=0}^{m=\infty}\,  m\,\exp(-m)\tau(m)}
{\sum_{m=0}^{m=\infty}\,\exp(-m)\tau(m)}-\frac{h}{\exp(h)-1}.
\end{align}

We now replace $m$ in the exponential by $\beta m$ and $h$ in the denominator of the second term by $\beta h$. This is immaterial if $\beta=1$, but this enables us to take derivatives with respect to $\beta$. Thus, the above expression becomes:
\begin{align}\label{uk}
U=\frac{\sum_{m=0}^{m=\infty} m \exp(-\beta m)\tau( m)}
{\sum_{m=0}^{m=\infty}\exp(-\beta m)\tau( m)}                                -\frac{h}{\exp(\beta\,h)-1}.
\end{align}

We then notice that:
\begin{align}\label{ahy}
U=-\frac{\partial \ln(Z)}{\partial \beta}\qquad Z(\beta,h)&=\p \exp(-\beta\,h) -1\q \sum_{m=0}^{m=\infty}  \exp(-\beta \,m)\tau(m).
\end{align}
Remember that we have employed a one-dimensional model so that some sources of energy related to the $x,y$ coordinates, corpuscle rotation and vibration, etc... are not accounted for. Note also that the summation in \eqref{ahy} is independent of $h$.

On the other hand, from \eqref{moy}
\begin{align}\label{moyg}
\ave{F}= \frac{1}{\exp(\beta\,h)-1}=\frac{1}{\beta}\frac{\partial \ln(Z)}{ \partial h}.
\end{align}

If we introduce the Helmholtz free-energy (the letter $A$ is from the German “Arbeit” or work): $A(\theta,h)\equiv -\theta\ln(Z(\theta,\,h))$ the expressions in \eqref{moy} and \eqref{ahy} are conveniently written:
\begin{align}\label{aab}
\ave{F}= -\frac{\partial A}{ \partial h}\qquad U= A-\theta \frac{\partial A}{ \partial \theta}.
\end{align}

From \eqref{aab} we obtain:
\begin{align}\label{bempty}
\delta Q&\equiv dU+\ave{F} \,dh=dA-\frac{\partial A}{ \partial \theta}d\theta-\frac{\partial A}{ \partial h}dh-\theta \,d\p\frac{\partial A}{ \partial \theta}\q=\theta \,dS\nonumber\\
 S&\equiv-\frac{\partial A}{ \partial \theta}=\ln(Z)+\theta\frac{\partial \ln(Z)}{\partial \theta}=\ln(Z)+\frac{\theta}{Z}\frac{\partial Z}{\partial \theta}=\ln(Z)-\frac{\beta}{Z}\frac{\partial Z}{\partial \beta},
\end{align}
where $\delta Q$ represents the heat delivered by the hot bath, from the law of conservation of energy. For any function $f(\theta,h)$ such as $U,\,A,\,S$: $df\equiv \frac{\partial f}{ \partial \theta}d\theta+\frac{\partial f}{ \partial h}dh$. Note that we employ only two independent variables, namely $\theta$ and $h$, so-that partial derivatives are un-ambigous. If the gas is in contact with a thermal bath ($\theta$=constant), $\delta Q$ is the heat gained by the bath. The quantity $S$ defined above is called \g{entropy}. In particular, if heat cannot go through the gas container wall (adiabatic transformation) we have $\delta Q=0$ that is, according to the above result: $dS=0$. Thus, in the context of the present work, adiabatic transformations are isentropic. After a cycle the entropy recovers its original value. Since the entropy does not change along adiabatic transitions we have: $S_l+S_h=0$. From \eqref{aab} we have therefore: $\beta_l Q_l+\beta_h Q_h=0$. From the law of conservation of energy: $Q_l+Q_h+W=0$, and we recover the Carnot efficiency: $\eta_C\equiv1-\beta_h/\beta_l$. We have therefore proven that $\theta$ is an absolute temperature.

Keeping $h$ a constant, it is straightforward to show from: $U=A-\theta dA/d\theta\Longrightarrow dU/d\theta=-\theta\, \partial^2A/\partial \theta^2$ and $S=-dA/d\theta\Longrightarrow dS/d\theta=- \partial^2A/\partial \theta^2$ that is: $dU/dS=(dU/d\theta)/(dS/d\theta)=\theta$. Thus, we may write $ \theta$ as: $\partial U(S,h)/\partial S$ if $U$ is considered a function of $S$ and $h$. This well-known expression, however, is not employed in the present work.

In the non-relativistic approximation ($\Longrightarrow \tau(Z)\propto \sqrt{Z}$): $\tau$ is proportional to $\sqrt{m}$. In the small $\beta$ (not-too-small temperature) limit, we may replace the summation in the expression of $Z$ in \eqref{ahy} by an integral, see \eqref{intn}:
\begin{align}\label{aabb}
\sum_{m=0}^{m=\infty}  \exp(-\beta \,m)\sqrt{m}\approx\int_{m=0}^{m=\infty} \,dm \exp(-\beta \,m)\sqrt{m}=\beta^{-3/2}\,\int_{0}^{\infty} dx \exp(-x)\sqrt{x}\propto \beta^{-3/2},
\end{align}
introducing $x\equiv \beta\,m$. The proportionality factor drops out in the derivation with respect to $\beta$, and needs not be given here. Accordingly, the internal energy in the non-relativistic and not-too-small temperatures approximation reads: 
\begin{align}\label{aabbb}
U=-\frac{\partial \ln(Z)}{\partial \beta}=-\frac{\partial \ln(\beta^{-3/2})}{\partial \beta}-\frac{h}{\exp(\beta\,h)-1}=\frac{3\theta}{2}-\frac{ h}{\exp(\beta\,h)-1}
\end{align}
For non-unity weights $w$, $h$ should be replaced by $w\,h$. In the absence of gravity: $\beta \,w\,h\to 0$, we finally obtain the well-known result:
\begin{align}\label{aabc}
U=\frac{3\theta}{2}-\theta=\frac{\theta}{2}\qquad \beta\,\theta\equiv 1.
\end{align}
The usual expression of the force $\ave{F}=\theta/h$ obtains in the absence of gravity, and not-too-small temperatures (Quantum effects are not accounted for). 

From \eqref{bempty} we see that if $Z$ is a product of two terms $Z_1\,Z_2$ as in \eqref{ahy} the entropy is the sum of two corresponding terms, say $S_1$ and $S_2$. For the first term: $Z_1(\beta,h)=\exp(-\beta\,h) -1$, we obtain: $S_1=\ln\left(\exp(-\beta h)-1\right)+\beta\,h/(1-\exp(\beta\,h) )\approx \ln(-h/\theta)$ if gravity is neglected ($\beta\,h\ll1)$. For the second term we have \eqref{aabb}: $Z_2\propto \beta^{-3/2}$, and to within un-important constants: $S_2=\ln(\theta^{3/2})$. It follows that the entropy $S=S_1+S_2$ is a function of $h^2\theta$ only, as we have obtained directly in \ref{otto} with $a$=1/2 as is appropriate to structureless corpuscles moving in one direction only.

In the limit presently considered, we recover straightforwardly the well-known Maxwell velocity distribution (in one dimension) and the definition of the absolute temperature as being proportional to the average corpuscle kinetic energy. Indeed, $U$ may be written as: $\theta/2+\Phi$, where $\Phi$ denotes the potential energy evaluated from the corpuscle density: $\rho(z)\propto \exp(-\beta z)$. In the limit where $w\,h\ll \theta$ the potential energy vanishes and the kinetic energy $K$ coincides with the total energy $U$. Thus $\theta$ is proportional to the average kinetic energy.  Since $\tau(E)\propto \sqrt{E}$, the energy distribution in the non-relativistic approximation is: $\propto \sqrt{E}\exp(-\beta E)$. This may be transformed into a probability distribution for $v$ with $E\propto v^2$ (remember that: $P_E(E) dE=P_v(v) dv$), leading to the usual Maxwell velocity distribution $\exp(-v^2)$ for one space dimension. For three space dimensions we have $v^2=v_x^2+v_y^2+v_z^2$, and the velocity distribution becomes: $P_v(v)\propto v^2 \exp(-v^2)$.

Remember that we have considered above evenly-spaced energies for mathematical convenience only. This is \emph{not} what is done in Quantum theory, where the \emph{actions} rather than the energies are evenly spaced. In the present situation, the corpuscle action is the area below the motion $z=z(t)$ for one period, since $w\,t$ and the momentum $p$ coincide.

For photons in three space dimensions:  $A =- \theta^4 h$ to within a constant factor. Then $F=\theta^4, ~U=3 \theta^4 h, ~S=4 \theta^3 h$. Another example is: $Z=\sum_k \exp(-\beta\eps_k)$, where the $\eps_k$ may depend on $h$. In that case: $S=-\sum_k p_k \ln(p_k),~p_k=\exp(-\beta\eps_k)/Z$.

\section{Air density as a function of altitude.}\label{plement}

It is well-known that as we climb on a mountain the air becomes less and less dense. Its composition may also change slightly because oxygen molecules are heavier that nitrogen molecules. We suppose that ideal pistons (defined are being weightless and able to move in the vertical direction without friction) may be removed without affecting the mechanical equilibrium. We obtain the change of air density with altitude and the force exerted by a corpuscle on a piston by employing an argument of consistency.

Let us consider superposed boxes, each of height $\Delta=1$, separated by weightless pistons. We suppose that when a box contains a single corpuscle, that corpuscle exerts a force unity on the upper piston and that the upper box contains a single corpuscle, as shown. Mechanical equilibrium of that piston requires a weight unity above the upper piston. The piston just below that one must support a weight 1+1=2, and thus that box must contain two corpuscles. The lower one must contain 1+1+2=4 corpuscles, and so on. It follows that if we label by $i=0$ the upper box, $i=1$ the box just below it, and so on, the number $N_i$ of corpuscles in the $i$-box must be equal to: 
\begin{align}\label{ag}
N_i=2^{i}
\end{align}
for the mechanical equilibrium to be achieved. The gas density is the number of corpuscles in a box since the boxes have the same unity height:
\begin{align}\label{akg}
\rho(z)\propto 2^{-z}.
\end{align} 
$\rho(z)$ is said to decay \emph{exponentially} as a function of the altitude $z=-i$. 

We have supposed above that the force exerted by a single corpuscle on the upper piston is unity when the box height $\Delta=1$. For consistency, the formula for an arbitrary $\Delta$ and $N$ corpuscles must be:
\begin{align}\label{mho}
F(\Delta)=\frac{N}{2^\Delta-1},
\end{align} 
which implies that $F(1)=1$ when $\Delta=1$ and $N=1$, in agreement with what has been said above. Notice that \eqref{mho} coincides with \eqref{moy} since $\Delta\equiv h$. Changing 2 into $e$ is presently immaterial.
 
\begin{figure}
\centering
\includegraphics[width=0.6\columnwidth]{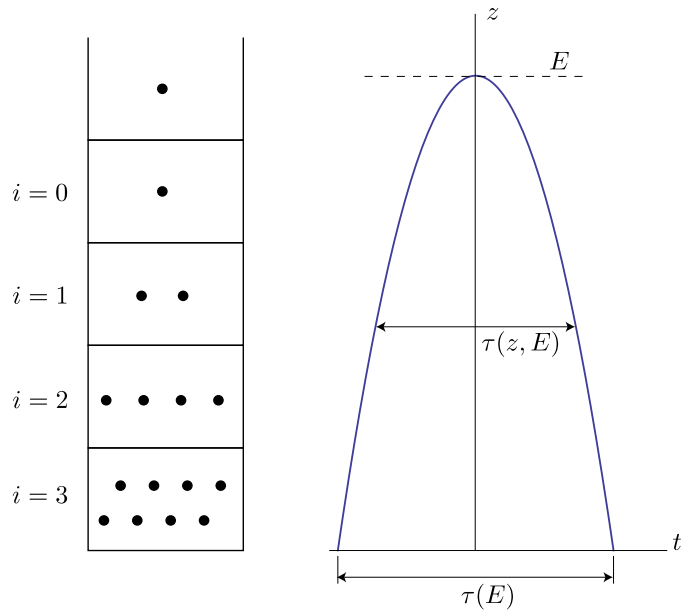}
\caption{This figure represents on the left a model of the atmosphere based on a concept of equilibrium. On the right, the curve shows that a single corpuscle trajectory would give an erroneous conclusion: an energy distribution is required.}
\label{atmos}
\end{figure}

Consider first an example. The number of corpuscles between $z=0$ and $z=-2$ is 3 as Fig.~\ref{atmos} shows. Since there are two boxes, the section height is $2\Delta=2$. The force on the upper piston is, according to the above formula: $\frac{3}{2^2-1}=1$ indeed. As another example, consider the three boxes from $z=0$ to $z=-3$. We have 7 corpuscles and the section height is $3\Delta=3$. The formula for the upper force is, according to \eqref{mho}: $\frac{7}{2^3-1}=1$. These examples suffice to show that the formulation is consistent. 

Alternatively, we may consider boxes of different heights. Consider two boxes separated by an ideal piston, each containing a single corpuscle of weight unity. The height of the upper section is $\Delta_1$ and the upper piston is supposed to support a weight $W$. If this is so, we have for the upper section of height $\Delta_1$: $W=F(\Delta_1)=\frac{1}{2^{\Delta_1}-1}$, or: $2^{\Delta_1}=\frac{1}{W}+1$. The lower section of height $\Delta_2$ must support a weight equal to the sum of the weight considered above and the weight unity of the upper section, that is, a weight $W+1$. It follows that: $W+1=\frac{1}{2^{\Delta_2}-1}$ implying that: $2^{\Delta_2}=\frac{1}{W+1}+1$. Accordingly: 
\begin{align}\label{mhol}
2^{\Delta_1+\Delta_2}-1=2^{\Delta_1}2^{\Delta_2}-1=(\frac{1}{W}+1)(\frac{1}{W+1}+1)-1=\frac{W+1}{W}\frac{W+2}{W+1}-1=\frac{2}{W}.
\end{align}

The two sections together contain two corpuscles and the total height is: $\Delta_1+\Delta_2$. Removing the piston separating sections 1 and 2 and applying the above general formula for $F$, we must have: 
\begin{align}\label{mjhol}
W=\frac{2}{2^{\Delta_1+\Delta_2}-1}=\frac{2}{2/W}=W,
\end{align}
which is indeed valid for any $W$-value. For three sections we have similarly:
\begin{align}\label{hol}
2^{\Delta_1+\Delta_2+\Delta_3}-1&=2^{\Delta_1}2^{\Delta_2}2^{\Delta_3}-1=(\frac{1}{W}+1)(\frac{1}{W+1}+1)(\frac{1}{W+2}+1)-1\nonumber\\
&=\frac{W+1}{W}~\frac{W+2}{W+1}~\frac{W+3}{W+2}-1=\frac{3}{W} \Longrightarrow W=\frac{3}{3/W}=W.
\end{align}
Generalization to any number of sections is straightforward.

If the corpuscle weight is $w$ instead of being unity, a more general formula for the force is: 
\begin{align}\label{mol}
F(\Delta)=\frac{w}{2^{w\Delta/\theta}-1},
\end{align}
where $\theta$, recognized earlier as being an absolute temperature, is here arbitrary. Since $w$ is a force and $\Delta$ a length, $w\Delta$ is an energy. It follows that $\theta$ has the dimension of an energy. When $\theta$ is divided by a standard energy $\kB= 1.38066...~  10^{-23}$ joules, called \g{Boltzmann constant}, we obtain a dimensionless quantity: the absolute temperature expressed in Kelvin, $T$. For historical reasons, $T$ is strictly defined as equal to 273.16 when the three phases (vapor, liquid, solid) of water (H-$^{16}$O-H) coexist. Then $\kB$ is the result of a measurement.

When the force of gravity (i.e., the weight $w$) is small, or more precisely when $w\,\Delta\ll\theta$, the approximation in \eqref{math} shows that the force is
\begin{align}\label{ol}
F\propto N \frac{\theta}{\Delta},
\end{align}
for $N$ corpuscles. The proportionality factor is unimportant because the temperature $\theta$ needs only be defined to within an arbitrary constant factor. We thus obtain the ideal gas law asserting that the product of the force exerted on the piston and the box height is proportional to the number of corpuscles and the temperature. 

In our one-dimensional model, the pressure $\textsf{P}$ corresponds to the average force $\ave{F}$, the volume $\textsf{V}$ to the height $h$, and $N=1$. Our result provides the ideal-gas law in a generalized form, taking into account gravity. In that case, the pressure varies as a function of altitude. More precisely, the force exerted by the corpuscle on the lower end of the cylinder exceeds the force exerted on the upper end (or piston) by the corpuscle weight. But in the absence of gravity, the forces exerted on both ends would be the same. We are introducing (static and uniform) gravity mainly because this helps clarify the concept of corpuscle energy: the corpuscle energy $E$ is the maximum altitude $z_m$ that the corpuscle would reach in the absence of the piston when the corpuscle weight is unity. The corpuscle bounces elastically off the ground and off the piston (if it reaches it) that is, without any loss or gain of energy. 

\paragraph{Arbitrary potential:}

We now consider the more general case where the weight depends on the $z$-coordinate. This is the case on earth since the weight is inversely proportional to the square of the distance $z+R$ from the earth center, where $R$ denotes the earth radius. Alternatively, one may say that the potential is: $\phi(z)\propto 1-\frac{1}{1+z/R}$. In the following, for mathematical convenience, we suppose that the weight varies by steps.

The motion of a corpuscle is then described  by an hamiltonien $E=H(z,p)$ which is the sum of a function of the momentum $p$ and a function of the altitude $z$: $H(z,p)=H(p)+\phi(z)$, where $\phi$ is the potential, not necessarily of the form: $w\,z$. The equations of motion are then: $dz/dt=v(p)=H'(p);\quad dp/dt=-w(z)\equiv\phi'(z)$, where $v(.),~w(.)$ are two nearly arbitrary functions and primes denote derivatives with respect to the argument. To help readers familiar with waves but not with the hamiltonian formalism, observe the correspondance: $E=\hbar \om, \quad p=\hbar k$, where $\hbar$ is a universal constant with the dimension of energy$\times$time, $\om=2\pi/time~period$ is the wave angular frequency, and $k=2\pi/space~period$ is called the wave number. Then $E=H(z,p)$ is the so-called dispersion equation. The first hamiltonian equation: $dz/dt=v=H'(p)$ then says that the group velocity $v$ is the derivative of $\om$ with respect to $k$. The second hamiltonian equation follows from the fact that in a time-independent system $\om$ is a constant of motion: $0=d\om=\frac{\partial \om}{\partial k}dk+\frac{\partial\om}{\partial z}dz\Longrightarrow \frac{dk}{dt}=-\frac{\partial\om}{\partial z}$.

When $w$ is a constant the solution is $z-z_o=-H(-w\,(t-t_o))/w$ as one can see by taking the derivative of that expression with respect to time, and $z_o,\,t_o$ are integration constants. A change of the corpuscle energy $E$ with respect to the $z=0$ level corresponds to a mere change of the $z_o$ constant, with the maximum altitude reached $z_m$ being equal to $E/w$. Omitting these constants, it follows from the previous relations that $\tilde{z}\equiv w\,z$ is a function of $\tilde{t}\equiv w\,t$ that does not depend on $w$. In the above discussion the function $v(p)$ is supposed to have been selected once for all, but is otherwise arbitrary (two exemples are: $H(p)=p^2/2$ and $H(p)=\sqrt{1+p^2}$). It follows that if we consider two weight values, say: $w_1$ and $w_2$, letting the subscripts 1 and 2 refer to the two weight values respectively, if $w_1\,t_1=w_2\,t_2$, then $w_1\,z_1=w_2\,z_2$. Accordingly $t_2(z_2)=\frac{w_1}{w_2}\,t_1(\frac{w_2}{w_1}\,z_2)$, where the times are considered functions of $z$ (rather than the opposite).

Let $\tau_i(Z)$ denote the round-trip time corresponding to a distance $Z$ from the top of the trajectory when the weight is a constant $w_i$ with $\tau_i(0)=0$ and, by convention, $\tau_i(Z)=0$ if $Z<0$. Setting $z_2=Z,\, t_1\to \tau_1, \, t_2\to \tau_2$ we have the relation: $\tau_2(Z)=\frac{w_1}{w_2} \tau_1(\frac{w_2}{w_1}Z)$. In particular, if the function $\tau(Z)$ denotes the round-trip time for a unity weight, we have for a weight $w_i$: $\tau_i(Z)=\frac{\tau(w_i\,Z)}{w_i} $. In the following we will only use this relation and the law of conservation of the energy along some corpuscle motion. 

We here consider weights that depend on the altitude $z$ by steps. Specifically, for mathematical convenience the weights are supposed to be: $1/\Delta_1$ for $0<z<\Delta_1\equiv z_1$, $1/\Delta_2$ for $\Delta_1<z<\Delta_1+\Delta_2\equiv z_2,\cdot \cdot \cdot$. It follows that the potential is incremented by 1 as one goes from one step to the next. The weight is a discontinuous function of $z$ but the potential is continuous. We are looking for the time spent by the corpuscle above $z_n$ for some integer $n$ as a function of the energy $E$. This time is obviously equal to zero if $E<n$ because the level $n$ is not reached. For higher energy values we have a sum of terms, each corresponding to the successive steps above $n$, that is:
\begin{align}\label{mdy}
0≤&E≤n & &\quad 0\nonumber\\
n≤&E≤n+1 & &\quad \tau_{n+1}(\frac{E-n}{w_{n+1}})=\Delta_{n+1} \tau(E-n)=\Delta_{n+1} d(E-n)\nonumber\\
n+1≤&E≤n+2 & &\quad \tau_{n+2}(\frac{E-n-1}{w_{n+2}})+\tau_{n+1}(\frac{E-n}{w_{n+1}})-\tau_{n+1}(\frac{E-n-1}{w_{n+1}}) \nonumber\\
& & &\qquad=\Delta_{n+2} \tau(E-n-1)+\Delta_{n+1} \tau(E-n)-\Delta_{n+1} \tau(E-n-1)\nonumber\\
& & &\qquad=\Delta_{n+1} d(E-n)+\Delta_{n+2} d(E-n-1)\nonumber\\
\dots\dots\dots&\dots\dots\dots &&\nonumber\\
n+i-1≤&E≤n+i & &\quad\sum_{k=1}^{i}\Delta_{n+k}\, d(E-n-k+1)
\end{align}
In the last expressions we have set: $\tau(E-i+1)-\tau(E-i)\equiv d(E-i+1)$, $i=1,2...$.

With an energy weight $\exp(-E/\theta)\equiv x^E$, the average round-trip time at $z_n$ is:
\begin{align}\label{ge}
\ave{T_n}&=\int_n^{n+1} dE\, x^E \Delta_{n+1} d(E-n)\nonumber\\
&+\int_{n+1}^{n+2} dE\, x^E \p\Delta_{n+1} d(E-n)+\Delta_{n+2} d(E-n-1)\q\nonumber\\
&+.........\nonumber\\
&=(\Delta_{n+1}x^n+\Delta_{n+2}x^{n+1}+\Delta_{n+3}x^{n+2}+\cdot \cdot \cdot)\,I(x)\qquad I(x)\equiv \int_0^{\infty} dE\, x^E  \,d(E).
\end{align}

Thus the probability that the corpuscle be above $z_n$ is:
\begin{align}\label{cdy}
\mathcal{B}(n)=\frac{\ave{T_n}}{\ave{T_0}}=\frac{\Delta_{n+1}x^{n}+\Delta_{n+2}x^{n+1}+\Delta_{n+3}x^{n+2}+...}{\Delta_{1}+\Delta_{2}x+\Delta_{3}x^2+...},
\end{align}
If we set the $\Delta$'s as unity we may identify $z$ and $n$ and obtain: $\mathcal{B}(z)=x^n=\exp(-z/\theta)$, that is the usual barometric law for a constant weight.

This expression may be compared with the result obtained traditionally from the ratio of the integrals of the generalized Boltzmann factor: $\exp(-\phi(z)/\theta)$ from $z_n$ to $\infty$ and from $0$ to $\infty$. Since $\phi(z_n)=n$, we obtain by that method for $z_{n-1}<z<z_{n}:$
\begin{align}\label{phi}
\int_{z_{n-1}}^{z_n} dz\,\exp \p-\frac{1}{\theta}( \frac{z-z_n}{\Delta_n}+n) \q\propto \Delta_{n} b^{n-1},
\end{align}
the proportionality factor being a function of $\theta$ only. This leads to our result for $\mathcal{B}(n)$ in \eqref{cdy}.

If the $\Delta_n$ are equal to 0 (corresponding to very large weights) except for $n=l$ and $n=h$, we find that the corpuscle can only be in the steps $l$ or $h$ of widths $\Delta_l\equiv g_l$, $\Delta_h\equiv g_h$ respectively. For simplicity we suppose that the \g{degeneracies} of these two levels are equal: $ g_l=g_h$. The ratio of the probability that the corpuscle be in $h$ and the probability that is be in $l$ is: $\exp\frac{\phi_l-\phi_h}{\theta}$. This is the generalized Boltzmann factor. In other words, if we consider only two levels with the same degeneracy $g$ we obtain that the ratio of the populations $\nu_l$, $\nu_h$ is given by the usual Boltzmann factor. In the two reservoirs heat engine shown in Fig.~\ref{deuxreservoirs} the population of the higher level is greater than the population of the lower level, a situation that can be achieved at equilibrium for a negative value of $\theta$, that is for a negative temperature. In laser theory this population inversion is considered to be a consequence of the pumping, an external source of energy.

\chapter{Conclusion}\label{conclu}

Some pre-socratic philosophers adopted a view-point that one may call \g{taking the side of things} or else, \g{emergentism}: the belief that the laws of thermodynamics emerge from the laws of corpuscle motion, the laws of living beings, including humans, from the laws of chemistry, and so on. This was the case in particular for Anaximander and Democritus. The main purpose of this work is to suggest that the fundamental results obtained by Carnot in 1824 relative to the operation of thermal engines, laws that can be viewed as being at the very foundation of most of physics,  could possibly have been obtained at the time of the ancient Greeks by pure reasoning, with perhaps the help of some crude observations, mainly for motivation. The present work may be used for pedagogical purposes.

We have been mostly concerned with the number of composition of integers with $g$ parts, and models involving only the potential energy, while in previous works authors mainly consider the kinetic energy, the absolute temperature being defined as being equal to the average kinetic energy. We have also considered the up and down motion of a corpuscle submitted to the earth gravity, employing a principle of simplicity, namely that the ideal gas law and the barometric law are \emph{independent} of the corpuscle laws of motion (non-relativistic, relativistic, or otherwise). The consideration of $z$-dependent corpuscle weights enables us to relate the various view-points presented, and in particular justifies the generalized Boltzmann factor.

\bibliographystyle{IEEEtran}
\bibliography{democrite}

\end{document}